\documentclass[12pt]{article}

\usepackage{fullpage}
\usepackage{epsf}
\usepackage{epsfig}
\usepackage{amsthm}
\usepackage{fancyhdr} % heading
\usepackage{amsfonts}
\usepackage{leftidx}
\usepackage{color}
\usepackage{amsmath}
\usepackage{setspace}
\usepackage{natbib}
\usepackage{hyperref}
\usepackage[ruled, vlined]{algorithm2e}
\usepackage{graphicx}
\usepackage{booktabs}
\usepackage{multirow}
\usepackage{multirow}
\usepackage{soul}

\begin{document}

\newcommand{\bm}[1]{\mbox{\boldmath $#1$}}
\newcommand{\mb}[1]{#1}%{\mathbf{#1}}
\newcommand{\bE}[0]{\mathbb{E}}
\newcommand{\bV}[0]{\mathbb{V}\mathrm{ar}}
\newcommand{\bP}[0]{\mathbb{P}}
\newcommand{\ve}[0]{\varepsilon}
\newcommand{\mN}[0]{\mathcal{N}}
\newcommand{\iidsim}[0]{\stackrel{\mathrm{iid}}{\sim}}
\newcommand{\NA}[0]{{\tt NA}}
\newcommand{\cB}{\mathcal{B}}
\newcommand{\R}{\mathbb{R}}
\newcommand{\Rp}{\R_+}

%%%%%%%%%%%%%%%%%%%%%%%%%%%%%%%%%%%%%%%%%%%%%%%%%%%%%%%%%%%%%%%%%%%%%%%%%%%%%%%%%%%%%%%%%%%%%%%%%%%%%%%%%%%

\title{\vspace{-1.5cm} Monotonic warpings for\\additive and deep Gaussian processes}
\author{
Steven D.~Barnett\thanks{Corresponding author \href{mailto:sdbarnett@vt.edu}{\tt sdbarnett@vt.edu}, Department of Statistics, Virginia Tech}
%\thanks{\em authors listed alphabetically}
\and Lauren J.~Beesley\thanks{Statistical Sciences Group, Los Alamos National Laboratory} 
\and Annie S.~Booth\thanks{Department of Statistics, Virginia Tech} \and
Robert B.~Gramacy\footnotemark[3] \and Dave Osthus\footnotemark[2]}

\maketitle

\vspace{-0.5cm}

\begin{abstract}

Gaussian processes (GPs) are canonical as surrogates for computer experiments
because they enjoy a degree of analytic tractability.  But that breaks when
the response surface is constrained, say to be monotonic. Here, we provide a
``mono-GP'' construction for a single input that is highly efficient even
though the calculations are non-analytic.  Key ingredients include
transformation of a reference process and elliptical slice sampling.  We then
show how mono-GP may be deployed effectively in two ways.  One is additive,
extending monotonicity to more inputs; the other is as a prior on injective
latent warping variables in a deep Gaussian process for (non-monotonic,
multi-input) nonstationary surrogate modeling. We provide illustrative and
benchmarking examples throughout, showing that our methods yield improved
performance over the state-of-the-art on examples from those two classes of
problems.

\end{abstract}

% \doublespacing % no double spacing for arXiv

\section{Introduction}
\label{sec:intro}

Gaussian processes (GPs) are popular as surrogate models for computer
simulation experiments \citep{santner2018design,gramacy2020surrogates}, as
general-purpose nonparametric regression and classification models in machine
learning \citep[ML;][]{williams2006gaussian}, and as priors for residual
spatial fields in geostatistics \citep{banerjee2004hierarchical}.  The reasons
are three-fold: (1) they furnish effective % not always accurate, hence DGPs etc.
nonlinear predictors with
well-calibrated uncertainty quantification (UQ); (2) they are nonparametric,
which means modeling fidelity increases organically with training data size
$n$; and (3) they have a high degree of analytic tractability, meaning that
they have closed expressions offloading numerics to linear algebra
subroutines. Efforts on the frontier of GP research involve: (1) scaling to
big $n$, circumventing a cubic bottleneck ($\mathcal{O}(n^3)$) in matrix
decomposition \citep[e.g.,][]{katzfuss2021general}; and (2) expanding fidelity
to circumvent stationarity \citep[e.g.,][]{paciorek2006spatial}.

This paper begins by taking things in the opposite direction: limiting GPs to
make the most of smaller $n$ from a constrained process.  However,
computational tractability and scalability will remain a theme throughout, and
we shall eventually turn to large-$n$ nonstationary modeling.  In particular,
we study GPs for modeling monotonic response surfaces. The literature here
largely focuses on a single input
\citep{tran2023monotonic,ustyuzhaninov2020monotonic,riihimaki2010gaussian,golchi2015monotone},
motivated by case studies in materials science and health care, but has
recently been extended to multiple inputs \citep{lopez2022high}.

Our idea involves undoing some of the analytic tractability that GPs are so
famous for, instead deploying a Markov chain Monte Carlo (MCMC) integration
technique known as elliptical slice sampling
\citep[ESS;][]{murray2010elliptical}.  ESS targets posterior inference for
(latent) variables under a multivariate normal (MVN) prior, which is the
essence of GP modeling.  While ESS cannot compete with analytic 
integration, it really shines with non-Gaussian response distributions like those
for classification (i.e., Bernoulli), where the integral is not analytic
and numerics are the only recourse.

In our setting the response distribution is still Gaussian, but we wish to
impose monotonicity on the underlying MVN latent field. Departing
from previous works, we think not in terms of constraints, implying certain
realizations must be (wastefully) rejected, but rather we create a process
which is monotonic by construction.  We do this first for one input
-- covering the vast majority of work on monotonic GPs
-- via transformation of the MVN process within ESS proposals. We then couple
   that with an additive structure \citep{lopez2022high} for a novel approach
   to monotonic modeling over multiple inputs.  

Our entire inferential apparatus is Bayesian, even for hyperparameters,
providing predictions with full UQ.  Generally, this would be intractable for
modest training data sizes, $n \ge 1000$ or so, even with analytic
integration.  Computational burdens are compounded with MCMC, even via ESS.  
So we make an
approximation.  Actually, we say ``approximation'' because what we do is a
coarsened version of the canonical GP on which it is based. However, we see
our setup as a different modeling apparatus altogether, much in the spirit of
inducing points
\citep[e.g.,][]{snelson2006sparse,banerjee2008gaussian,cole2021locally} or
Vecchia \citep[e.g.,][]{katzfuss2021general,banerjee2004hierarchical}
approximations.  We establish a low-dimensional grid-based reference process,
i.e., for $n_g \ll n$ quantities, linking individual inputs to an output,
possibly under a monotonic transformation, which limits cubic bottlenecks to
$\mathcal{O}(n_g^3)$.  Rather than inducing a low-rank structure for the
entire spatial field, we use simple linear interpolation.  Unlike
Vecchia, we do not need to choose a data ordering or neighborhood set.
% We show how to extend the low dimensional reference processes to
% monotonic GPs in higher input dimension through an additive model.

We then pivot to another application, involving injective priors over the
warpings provided by the latent layers of a deep Gaussian process
\citep[DGP;][]{damianou2013deep,sauer2023active,sauer2023vecchia}. DGPs are a
recently popular nonstationary modeling apparatus in ML and computer
experiments \citep{sauer2023non}.  
% Actually, DGP warpings were the original motivation for our mono-GP.  
We wish to address a DGP-related concern recently raised in spatial and
geostatistical settings \citep{zammit2022deep}.  While a more cavalier
attitude -- ``let the posterior do whatever it wants'' -- can sometimes be
advantageous, we agree with \citeauthor{zammit2022deep}~that non-injective
warpings are less interpretable and, in some (possibly most) cases, lead to
over-fitting and inferior results out-of-sample. We show that applying our
mono-GP separately to the warping of each individual input of a deep GP
(mw-DGP), which guarantees an injective input map by construction, is both
easier to interpret than an unconstrained warping and leads to more accurate
predictions for surrogate modeling of computer simulation experiments. % We get the best results when
% nonstationary dynamics are axis-aligned, which is a class of problems
% previously dominated by the treed Gaussian process
% \citep{gramacy2008bayesian}.

Toward that end, the paper is outlined as follows.  We review GPs and ESS in
Section \ref{sec:review}, ending with a novel reference process idea.  Section \ref{sec:mono} introduces a monotonic
transformation for one input, extended additively for multiple inputs in
Section \ref{sec:add}. Section \ref{sec:mwDGP} discusses monotonic warpings for
DGPs.  Illustrations and empirical benchmarking are provided throughout.
Section \ref{sec:discuss} concludes with a discussion. Code reproducing all
examples is provided at \url{https://bitbucket.org/gramacylab/deepgp-ex/},
via {\tt deepgp} on CRAN \citep{deepgp}.

\section{Basic elements}
\label{sec:review}

There are several, more-or-less equally good ways to formulate a GP for
regression.  Here we have chosen one that is well-suited to our narrative.
For review with an ML perspective, see \cite{williams2006gaussian}. For
computer experiments, see \citet{santner2018design}.  The setup here closely
mirrors \citet[][Section 5.3]{gramacy2020surrogates}.  After laying out the
basics, we propose an alternative course for inference where the main integral
is performed numerically.  This allows us to introduce a new approximation
which is particularly handy for mono-GP.

\subsection{Canonical GP formulation and inference}
\label{sec:gp}

Suppose we wish to model data pairs $(x_1, y_1), \dots, (x_n, y_n)$
where $x_i \in \mathbb{R}^p$ and $y_i \in \mathbb{R}$ in a regression setting,
i.e., where $y_i = \mu + \nu f(x_i) + \varepsilon_i$ and $\varepsilon_i
\stackrel{\mathrm{iid}}{\sim} \mathcal{N}(0, \sigma^2)$, $i=1,\dots,n$. Scalar
$\mu$ and $\nu$, representing mean and scale, are sometimes called
hyperparameters in this context because their settings represent more of a
fine-tuning. The main inferential object of interest is the unknown function
$f:\mathbb{R}^p \rightarrow \mathbb{R}$, and depending on how it is modeled, 
values of $(\mu, \nu) = (0,1)$ may suffice.  In fact, many authors use
that simplification, pushing notions of center and amplitude onto  $f$ or to
user pre-processing.  We introduce them here, separate from $f$, for
compatibility with some of our downstream modeling choices.

Placing a GP prior on $f$ amounts to specifying that any joint collection of
realizations of $f$, say at the $n$ inputs $x_i$, is MVN, i.e., $F_n \sim
\mathcal{N}_n(0_n, C_n)$ where $0_n$ is an $n$-vector of zeros, and $C_n$ is
an $n \times n$ positive definite correlation matrix. Usually $C_n^{ij} =
\exp\{ - \sum_{k=1}^p (x_{ik} - x_{jk})^2/\theta_k\}$ for positive {\em
lengthscale} hyperparameters $\theta = (\theta_1, \dots, \theta_p)$.  There
are many variations on the details of this construction, particularly
involving the so-called {\em kernel} determining $C_n$, but they are not
material to our discussion.  For now, fix particular values for all
hyperparameters $(\mu, \nu, \sigma^2, \theta)$ so that the focus is $F_n$.
Summarizing, for row-stacked $X_n$, we have
\begin{align}
\mbox{Likelihood:} && Y_n \mid X_n, F_n  &\sim \mathcal{N}_n(\mu + \nu F_n, \mathbb{I}_n \sigma^2), \label{eq:model} \\
\mbox{and Prior:} && F_n \mid X_n &\sim \mathcal{N}_n(0_n, C_n) & \mbox{ where } C_n &\mbox{ depends on $X_n$ and $\theta$.} \nonumber
\end{align}
Bayes' rule provides
\begin{equation}
p(F_n \mid X_n, Y_n) = \frac{p(Y_n \mid F_n, X_n) \cdot p(F_n \mid X_n)}{p(Y_n \mid X_n)}. \label{eq:bayes}
\end{equation}
The denominator $p(Y_n \mid X_n)$ is sometimes called a {\em marginal likelihood} because it may be evaluated by integrating over the likelihood
\begin{equation}
p(Y_n \mid X_n) = \int_{\mathbb{R}^n} p(Y_n \mid f, X_n) \cdot p(f \mid X_n) \; df. \label{eq:marg}
\end{equation}
In fact, this quantity has a closed form expression so that inference for
$F_n$ can be reduced to linear algebra.   The marginal likelihood can also be
used to learn settings for hyperparameters, either via maximization or
posterior integration. Notice that the dimension $n$ of $F_n$ plays an
important role.  We may also integrate over the posterior for $f(x)$ at new
predictive locations $x$ in closed form. We don't provide these equations;
they can be found in nearly every GP reference.  Suffice it to say that the
predictive distribution is Gaussian for $x$, or jointly MVN for a collection
of $n'$ locations $\mathcal{X} \in \mathbb{R}^{n' \times p}$, 
conditional on hyperparameters.

\subsection{Elliptical slice sampling}

An alternative way to perform posterior inference for $F_n$ via
Eq.~(\ref{eq:bayes}), say if you were unaware of textbook results, would be
via Markov chain Monte Carlo (MCMC).  In so doing, you would approximate the
marginal likelihood (\ref{eq:marg}), with accuracy depending on myriad factors
including your choice of MCMC algorithm and computational effort.  A
particularly efficient choice -- but not better than jumping right to the
closed form expression -- is elliptical slice sampling
\citep[ESS;][]{murray2010elliptical}.  ESS is ideally suited to our situation:
MVN prior in high dimension (large $n$) with simple likelihood
(\ref{eq:model}), which for us is iid Gaussian.

\begin{figure}[ht!]
\centering
\includegraphics[scale=0.41,trim=0 50 0 30]{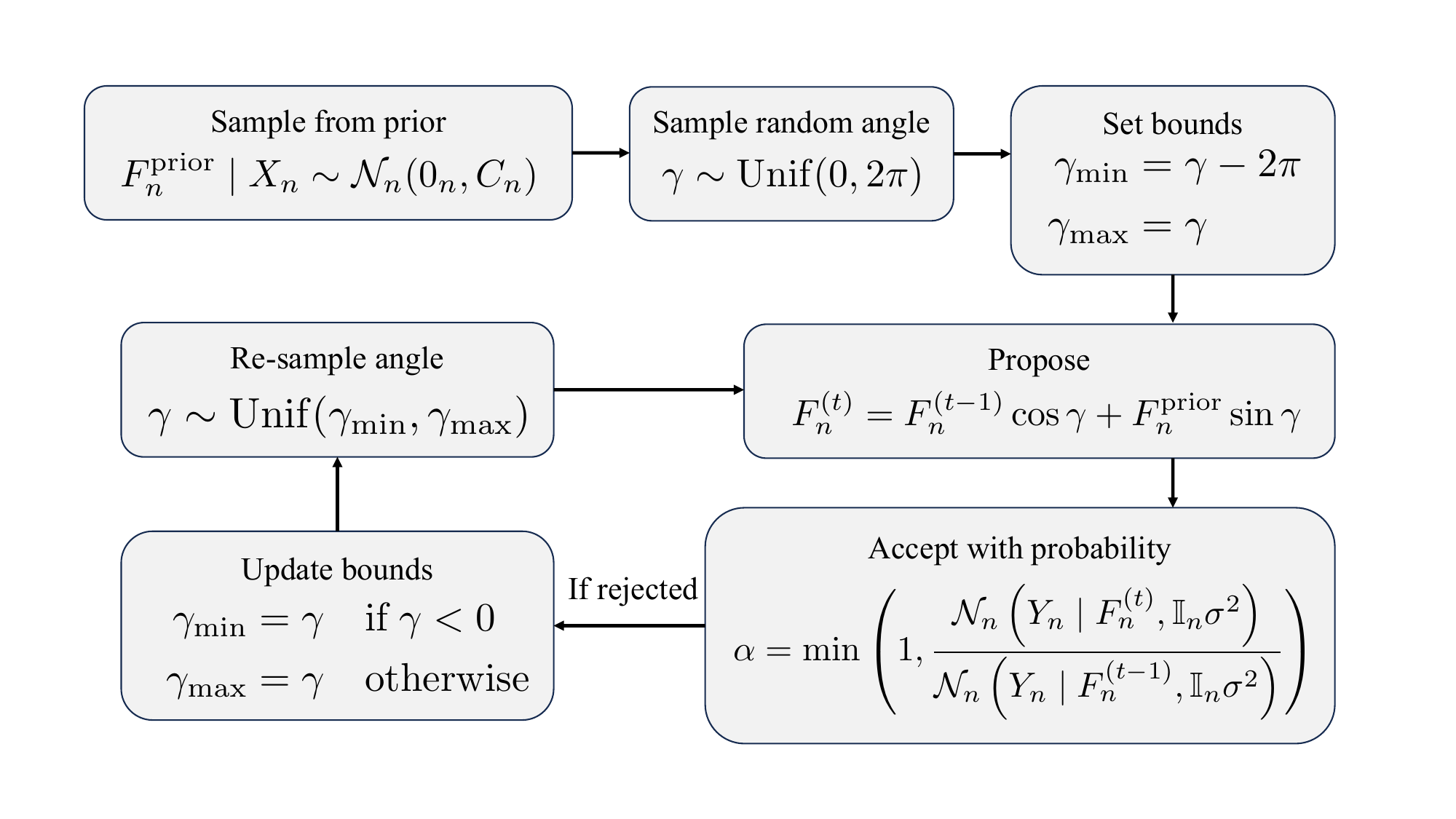}
\caption{Diagram of one iteration of the ESS algorithm.
\label{f:ess}}
\end{figure}

ESS was not created for ordinary GP regression, because it is unnecessary there, but
instead for models like classification where the MVN prior is the same as ours
but the likelihood is Bernoulli after a logit transformation.  In that context
the $F_n$ are latent and explicit (numerical integration is required). Perhaps
the most attractive features of ESS are that there are no tuning parameters,
and it is rejection-free in the sense that MCMC sample $t$ automatically
iterates/adjusts from $F^{(t-1)}_n$ so a novel $F_n^{(t)}$ may be returned.  The
algorithm is depicted diagrammatically in Figure \ref{f:ess}.
It is important to recognize where the work is being done, computationally
speaking.  Everything except sampling from the prior (top-left) is at worst
$\mathcal{O}(n)$ flops, whereas generating $F_n^{\mathrm{prior}}$ is
$\mathcal{O}(n^3)$ via Cholesky  \citep[][Appendix B]{gelman1995}. MCMC
mixing, that is the diversity of $F_n^{(t)}$ relative to $F_n^{(t-1)}$, is
generally excellent and often it only takes a few iterations to ``accept'' in
the likelihood ratio (bottom-right), after which we update $t \leftarrow
t+1$ and move on to the next sample.

For prediction, rather than $F_n^{\mathrm{prior}} \mid X_n \sim \mathcal{N}_n(0_n, C_n)$ 
in the first step, instead draw 
\begin{equation}
\left(\begin{array}{c} F_n^{\mathrm{prior}} \\ 
\mathcal{F}_{n'}^{\mathrm{prior}}
\end{array}\right) \sim \mathcal{N}_{n + n'}
\left(
\left[\begin{array}{c} 0_n \\ 0_{n'} \end{array} \right],
\left[\begin{array}{cc} C_n & C_{nn'} \\ 
C_{n'n}  & C_{n'} \end{array} \right]
 \right),
 \label{eq:pred}
\end{equation}
where $C_{n'}$ is defined similarly to $C_n$ except with inputs
$\mathcal{X}_{n'}$ and $C_{nn'}^{ij}$ involves a kernel calculation on
the $i^\mathrm{th}$ row of $X_n$ and the $j^\mathrm{th}$ row of
$\mathcal{X}_{n'}$.  Then, only use $F_n^{\mathrm{prior}}$ in subsequent
calculations for ESS following the diagram in Figure \ref{f:ess}.  When
eventually an acceptable combination of previous/prior sample is ``accepted'',
return $\mathcal{F}_{n'}^{(t)} =
\mathcal{F}_{n'}^{(t-1)} \cos \gamma + \mathcal{F}_{n'}^{\mathrm{prior}} \sin
\gamma$ as the next sample of the unknown function at $\mathcal{X}_{n'}$.

\subsubsection*{An illustration}

\begin{figure}[ht!]
\centering
\includegraphics[scale=0.6, trim=0 15 15 5]{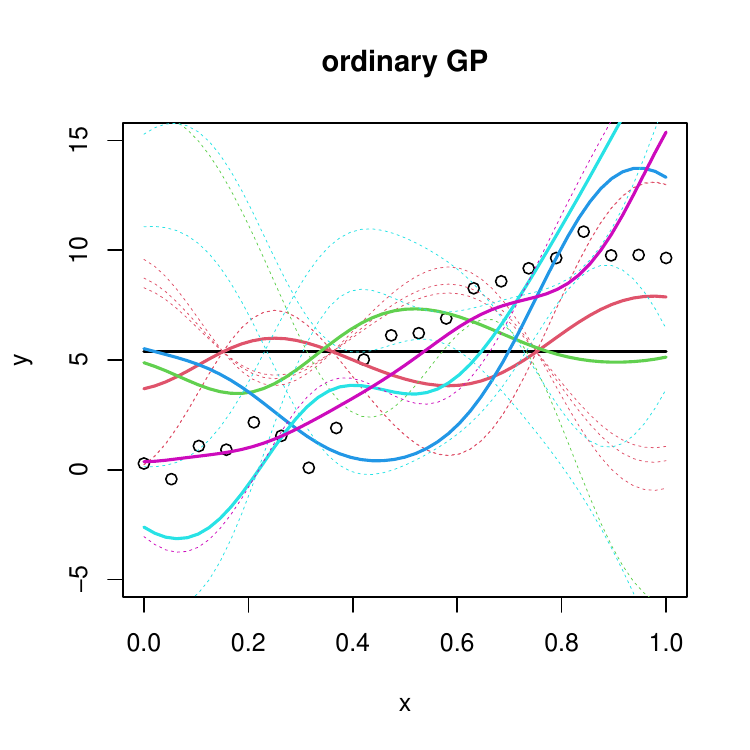}
\includegraphics[scale=0.6, trim=50 15 0 5,clip=TRUE]{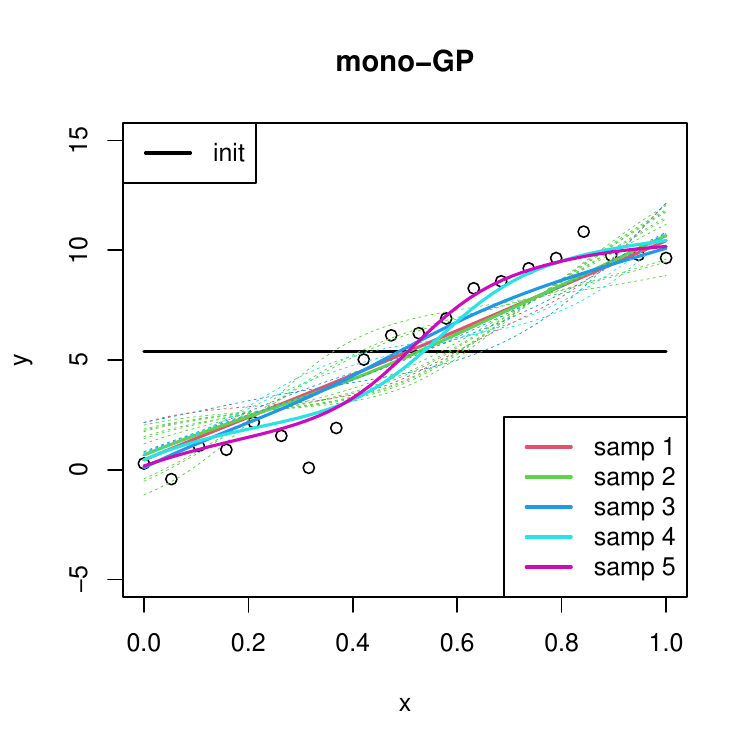}
\caption{ESS in 1d.  Thick solid lines represent
``accepted'' ESS samples $F_n^{(t)} \mid F_n^{(t-1)}$, beginning with
$F_n^{(0)} = 5$.  Thinner, dashed lines of the same color represent
rejected proposals.  
%The {\em left} panel corresponds to an
%ordinary GP, whereas the {\em right} panel is a monotonic GP, discussed
%later in Section \ref{sec:mono}.
\label{f:essviz}}
\end{figure}

To illustrate, consider a simple logistic response in one input dimension
observed with a small amount of noise at $n=20$ equally-spaced inputs $X_n$.
The left panel in Figure \ref{f:essviz} shows the outcome of five samples via
ESS starting with an initial $F_n^{(0)} \approx 5$, i.e., constant, indicating
no relationship between $x$ and $y$.  Ignore the right panel for now. That,
and other details for this visual are deferred to Section \ref{sec:monopost}
so as not to detract from the focus on $F_n$.

Each thick colored line represents a subsequent $F_n^{(t)}$ for some $t$.
Thinner, dashed lines of identical color indicate rejected samples within ESS
before the thick one was accepted.  Observe that it doesn't take
many iterations before one is accepted, and that the chain converges very
quickly to something reasonable from a disadvantageous
starting position.

Actually, the lines in Figure \ref{f:essviz} are not samples $F_n^{(t)}$,
indexed by just $n=20$ inputs $X_n$, but predictive ones $\mathcal{F}_n^{(t)}$
on a denser grid $\mathcal{X}_{n'}$, as described around Eq.~(\ref{eq:pred}).
The reason for this is two-fold.  One is that with just $n=20$ coordinates,
the plotted $F_n^{(t)}$ wouldn't look smooth to the eye.  You'd see nineteen
small, connected line segments.  By using a slightly larger grid
$\mathcal{X}_{n'}$ with $n'=50$, we get line segments (which are technically
still there, because that's how {\sf R} draws curves from points) that look
like samples of a smooth function. % intentionally merged these two paragraphs
The second reason is the converse of the first:  a collection of fifty
responses on a grid of inputs is enough to convey all relevant information
about a function of one input, at least from the perspective of visual
perception.  This leads us to our first novel idea.

\subsection{Reference process for one input}
\label{sec:ref}

Consider, as in the illustration above, a single input ($p=1$) coded to the
unit interval: $X_n \in [0,1]^n$.  Now, create an evenly-spaced grid  % if we don't say
% evenly-spaced here, then we don't really say it anywhere
$X_g$ in $[0,1]$, akin
to knots or inducing points \citep{snelson2006sparse,banerjee2008gaussian}
with corresponding $F_g$. There is really no difference between the latent
process at $X_g$ and the one at $\mathcal{X}_{n'}$ from the illustration above
except that $F_g$ is on a pre-determined grid, whereas predictive
locations of interest could be anywhere and of any size.  In fact, we prefer a
grid of size $n_g = 50$, though ultimately this is a tuning parameter trading
off computational effort and resolution.  [See Appendix \ref{app:prior}.]

Now, an inducing points approach uses GP conditioning to relate $(X_g, F_g)$
to training $(X_n, Y_n)$ or testing $(\mathcal{X}_{n'}, \mathcal{Y}_{n'})$.
This is coherent from a modeling perspective, but comes at a cubic expense in
all sizes: $\mathcal{O}(n^3 + {n'}^3 + n_g^3)$.  Plotting in {\sf R} could
work that way too, but it's clearly overkill.  For most situations, linear
interpolation from $(X_g, F_g)$ to $X_n$ or  $\mathcal{X}_{n'}$ works great:
you can't tell it's not a curve with modest $n_g$ and the computational cost is
greatly reduced: $\mathcal{O}(n_g^3 + n + n')$.  So this is what we propose to
do, but inside MCMC posterior sampling via ESS.  Everything is performed on
the reference process $(X_g, F_g)$, but whenever we need to evaluate a
likelihood for $Y_n$ or a prediction for $\mathcal{X}_{n'}$ we linearly
interpolate to $F_n$ or $\mathcal{F}_{n'}$. Modifications to Figure
\ref{f:ess} are as follows: keep track of $F_g^{(t-1)}$, propose
$F_g^{\mathrm{prior}}
\mid X_g$ (top-left), form $F_g^{(t)} = F_g^{(t-1)} \cos \gamma +
F_g^{\mathrm{prior}} \sin \gamma$, but to calculate the acceptance probability
(bottom-right) first interpolate to $F_n^{(t)}$ and $F_n^{(t-1)}$ before
evaluating likelihoods.

Actually, you can think about the linear interpolation as being inside the
likelihood itself.  In this way, it is not an approximation but a totally
different model.  The part of it that is a GP, i.e., the MVN prior on latent
functions, is unchanged: for any inputs $X$ whether that is $X_g$, $X_n$ or
$\mathcal{X}_{n'}$, the prior is MVN: $F \sim \mathcal{N}(0, C(X))$, abusing
notation somewhat. But when evaluating the likelihood (\ref{eq:model}) we
first ``snap'' from the reference process to the actual one: $F_n =
\mathrm{LI}(X_n, X_g, F_g)$, where LI stands generically for linear
interpolation, say via {\tt approx} in {\sf R}, but any fast interpolator will
do.  We provide the details of our own in Appendix \ref{app:fo_approx}.  This
is completed with a prior over the {\em reference} latent process $F_g \mid
X_g \sim \mathcal{N}_{n_g}(0_{n_g}, C_g)$,  where $C_g$ depends on $X_g$ and
$\theta$. We shall write out a full hierarchical model, extending
Eq.~(\ref{eq:model}), once we add the monotonic ingredient in Section
\ref{sec:mono}.  Although the $F_g \mid X_g$ can
have a coarsening effect with small $n_g$ [as we explore in
Appendix \ref{app:approx}], the model for $Y_n$ still uses all data to learn
the latent function and hyperparameters, again delayed until Section
\ref{sec:mono}.

We limit our application of this idea to a single input.  It may be possible
to extend to $p \geq 2$, but we doubt its practicality. Gridding out $[0,1]^p$
with the same density as in 1d (e.g., $n_g = 50)$ would require $n_g = 50^p$
via Cartesian product.  Even an un-gridded space-filling set of knots $X_g$
would still require large $n_g$ to fill the input volume, obliterating any
potential computational savings while severely coarsening the reference
process.  Note that inducing points also present issues in modest $p$ without
large $n_g$ \citep[e.g.,][]{wu2022variational}. Nevertheless, we find a 1d
reference process useful for a class of monotonic additive models [Section
\ref{sec:add}] and as priors on injective input warpings for DGPs [Section
\ref{sec:mwDGP}].  In those settings it is not just a computational
convenience, but a crux of the whole enterprise.

\section{Monotonic Gaussian processes}
\label{sec:mono}

Here we extend the reference process of Section \ref{sec:ref} to
monotonic functions.  

%Although restrictive, this is an important building
%block to higher-powered applications in later sections. 

\subsection{Non-decreasing transformation}
\label{sec:mono1} 

Consider again $p=1$ and $X_n \in [0,1]^n$ but now additionally suppose
$X_n$ is ordered: $x_{i} \leq x_{i+1}$. Now, we change
notation to refer to the Gaussian latent process as $Z$ rather than $F$. That
is, let $Z_n \sim
\mathcal{N}_{n}(0_n, C_n)$.  A transformation of $Z$ is designed so that $F$
is monotonic in $X$.  Many transformations could work well; see commentary in Appendix \ref{app:prior}. Our preferred transformation involves 
three steps: (1) exponentiate for positivity;  (2) cumulatively sum so
non-decreasing; (3) normalize to $[0,1]$.  In formulas, if $Z^{(0)} \sim
\mathcal{N}(0, C)$, then
\begin{align}
Z^{(1)} &= \exp{Z^{(0)}}, & 
z^{(2)}_i &= \sum_{j=1}^{i} z^{(1)}_j, & 
f_i \equiv z^{(3)}_i &= \frac{z^{(2)}_i - \min(Z^{(2)})}{\max(Z^{(2)}) - \min(Z^{(2)})}, \label{eq:mono}
\end{align}
for $i=1,\dots,n$.
\begin{figure}[ht!]
\centering
\includegraphics[scale=0.6, trim=0 15 15 50,clip=TRUE]{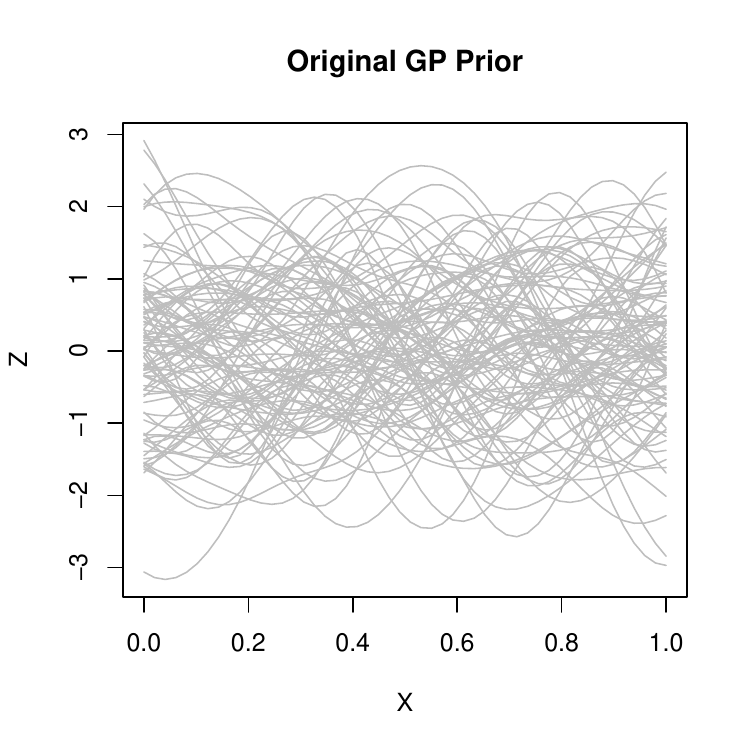}
\includegraphics[scale=0.6, trim=0 15 0 50,clip=TRUE]{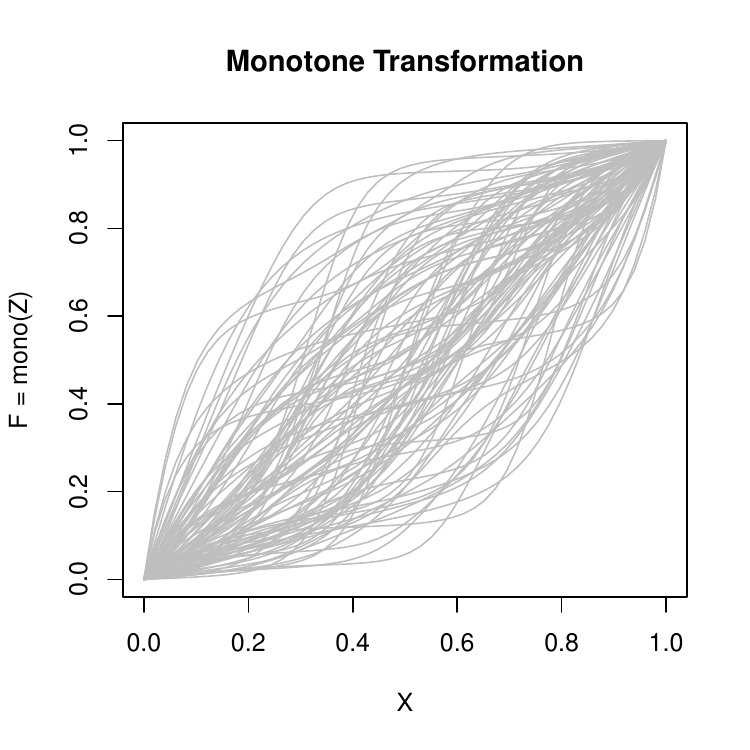}
\caption{Converting 100 samples from GP prior (left) into monotonic
realizations (right). \label{f:prior}}
\end{figure}
Figure \ref{f:prior} provides a visual.  On the left are ordinary $Z$ from a
GP using a lengthscale of $\theta = 0.01$, and on the right their
monotonically transformed (\ref{eq:mono}) counterparts $F$. The cumulative sum
(after ensuring positivity) provides the key non-decreasing ingredient.
Post-processing to unity is not essential, but helps inferential details
described later.

Yet the cumulative sum is potentially problematic in the context of
prediction, which is our main objective. Suppose we have training data at
$X_n$ and testing locations $\mathcal{X}_{n'}$, also ordered. We could apply
Eq.~(\ref{eq:mono}) to $Z = (Z_n, \mathcal{Z}_{n'})$ drawn as in
Eq.~(\ref{eq:pred}), but with $Z$s instead of $F$s.  Yet we would get
different $F_n$-values from Eq.~(\ref{eq:mono}) depending on whether we
transformed $\mathcal{Z}_{n'}$ along with the $Z_n$ values, and also on which
$\mathcal{X}_{n'}$ are used.
\begin{figure}[ht!]
\centering
\includegraphics[scale=0.6, trim=0 15 15 50,clip=TRUE]{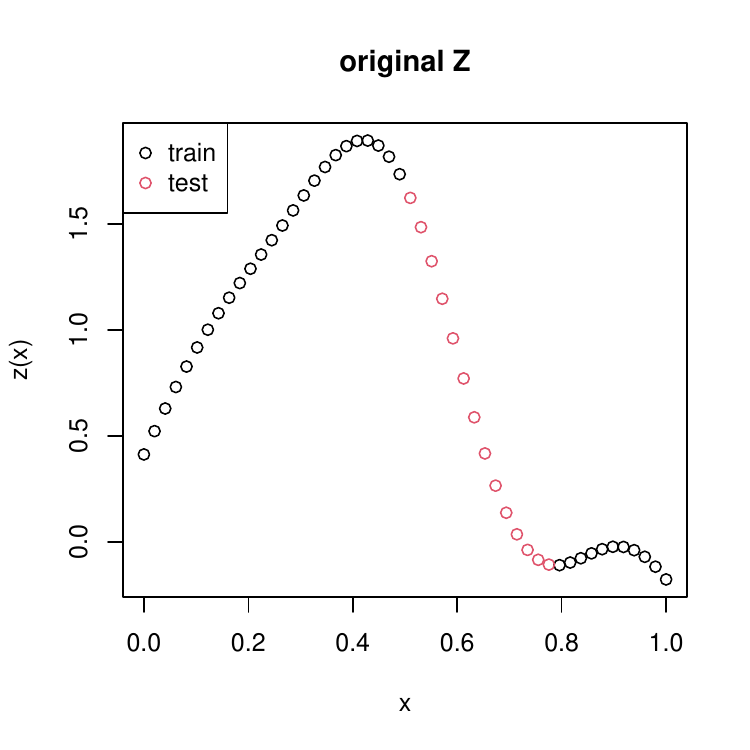}
\includegraphics[scale=0.6, trim=0 15 0 50,clip=TRUE]{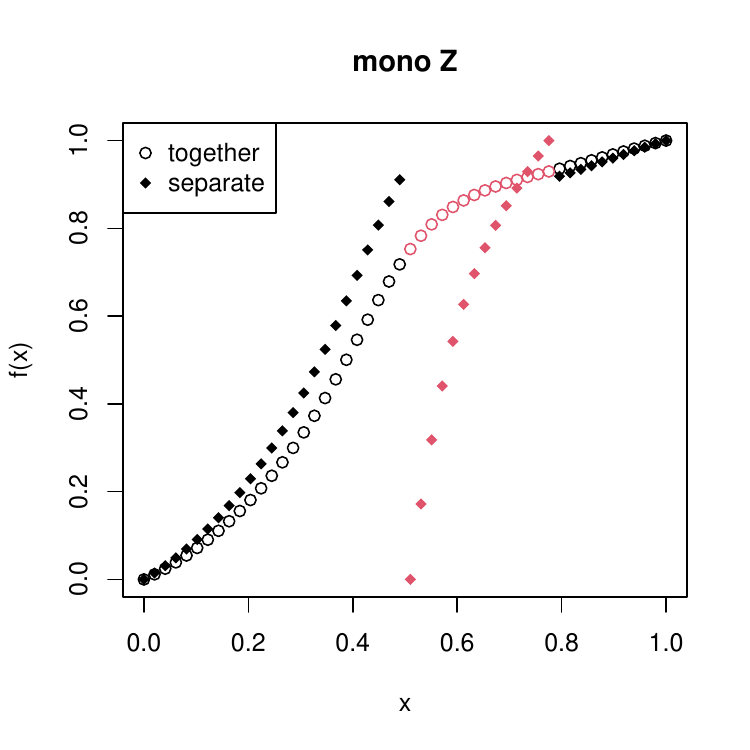}
\caption{Joint or separated mapping of training and testing $Z$s (left) into $F$s (right).
\label{f:problem}}
\end{figure}
For example, the left panel of Figure \ref{f:problem} shows a single
realization of combined training and testing $Z$-values. Notice that the
monotonic $F$-values you get on the right depend on whether you apply
Eq.~(\ref{eq:mono}) jointly (together for train and test at once) or separately.
Also observe that when $\mathcal{X}_{n'}$ is heavily concentrated in one part
of the input space relative to $X_n$ you get a discontinuity in $F_n$. We want
smooth latent functions, and more broadly we don't want a model for latent
$F_n$ that is affected by where you're trying to predict $\mathcal{X}_{n'}$.
$F_n$ should depend only on $X_n$.

The reference process that we introduced in Section \ref{sec:ref} can fix both
things, whereby everything is anchored to one collection $(X_g, Z_g)$ via
linear interpolation and transformation. Eq.~(\ref{eq:model}) is unchanged; we
still have $F_n = \mathrm{LI}(X_n, X_g, F_g)$.  For prediction we may
similarly calculate $\mathcal{F}_{n'} = \mathrm{LI}(\mathcal{X}_{n'}, X_g,
F_g)$, and in so doing we prevent a choice of predictive locations from
affecting training quantities through the cumulative sum.   As a bonus, the
reference process enjoys all of the computational advantages boasted for
ordinary GPs earlier. Note only the reference $X_g$ must be ordered in this
setup, not $X_n$ or $\mathcal{X}_{n'}$.

\subsection{Full model and posterior inference}
\label{sec:monopost}

Our full mono-GP hierarchical model is provided below. 
\begin{align}
\mbox{Likelihood:} && Y_n \mid X_n, F_n  &\sim \mathcal{N}_n(\mu + \nu (F_n - \textstyle\frac{1}{2}),
\mathbb{I}_n \sigma^2) 
&& \mbox{notice the } - \textstyle\frac{1}{2} \label{eq:monomodel} \\
 && F_n &= \mathrm{LI}(X_n, X_g, F_g) 
 && \mbox{say via {\tt approx} in {\sf R}} \nonumber \\
 && F_g &= \mbox{\tt mono}(X_g, Z_g) && \mbox{following Eq.~(\ref{eq:mono})}
 \nonumber \\
\mbox{and Priors:} && Z_g \mid X_g &\sim \mathcal{N}_{n_g}(0_{n_g}, C_g) 
&& C_g \mbox{ depends on $X_g$ and $\theta$} \nonumber \\
&& p(\mu, \sigma^2) &\propto 1/\sigma^2 && \mbox{improper Jeffreys prior} \nonumber \\
&& \nu &\sim \mathrm{Gamma}(\alpha_\nu, \beta_\nu) && \mbox{diffuse, e.g., $\alpha_\nu=\beta_\nu=10^{-3}$} \nonumber \\
&& \theta &\sim \mathrm{Gamma}(\alpha_\theta, \beta_\theta) \nonumber
&& \mbox{e.g., $(\alpha_\theta, \beta_\theta) = (3/2, 5)$}
\end{align}
Some quick commentary is in order.  Notice how we center $F_n$ around zero by
subtracting $1/2$.  This is because we normalized to $[0,1]$ as part of
Eq.~(\ref{eq:mono}), and we wish to interpret $\mu$ as a ``centering'' location
parameter.  But it is not essential; we can skip it or instead normalize
directly to $[-0.5, 0.5]$.  Our choices here are motivated by a compromise
between modeling goals in Section \ref{sec:mono}--\ref{sec:add} and our deep
GP application in Section \ref{sec:mwDGP}.

Linear interpolation (LI) and the monotonic transformation ({\tt mono}) could be
composed into a single step, as $\mathrm{LI}(X_n, X_g, \mathrm{mono}(X_g,
Z_g))
\equiv \mbox{\tt monoref}(X_n, X_g, Z_g)$, and this is how it is in our
implementation.  We never need to do one without the other, but we thought it
helpful to have them separate in Eq.~(\ref{eq:monomodel}).  Both are
completely deterministic, working together to define a GP prior on the latent
random function $f(x)$. In our posterior sampling scheme described
momentarily, we only save $Z_g^{(t)}$-values as everything else we need, like
$F_n^{(t)}$ and $\mathcal{F}_{n'}^{(t)}$, can be derived from these through
{\tt monoref}.

A Jeffereys prior for location-scale models \citep[see,
e.g.,][]{hoff2009first} yields a proper posterior for $n>1$ and may be
upgraded to Normal-Inv-Gamma if desired.  Both are conditionally conjugate for
the Gaussian likelihood, allowing $\mu$ and $\sigma^2$ to be integrated out.
Let
\begin{align}
\hat{\mu} &= \frac{1}{n} \sum_{i=1}^n y_i - \nu(f_i - \textstyle \frac{1}{2}) 
&& \mbox{ and } &
s^2 &= \frac{1}{n-1} \sum_{i=1}^n (y_i - \hat{\mu} - \nu(f_i - \textstyle\frac{1}{2}))^2 \label{eq:summ}
\end{align}
denote summary statistics for residual location and scale.  A marginal
likelihood follows
\begin{equation}
p(Y_n \mid F_n, \nu) = \int\!\!\int \mathcal{N}_n(Y_n \mid \mu + \nu (F_n - \textstyle\frac{1}{2}), \mathbb{I}_n \sigma^2) \frac{1}{\sigma^2} \,
d\mu d\sigma^2 \propto \left(\frac{(n-1)s^2}{2}\right)^{-\frac{n-1}{2}}.
\end{equation}
This is a relatively standard result, so we don't provide derivation details.

What's non-standard is its use within ESS sampling for $F_n$ and Metropolis
for $\nu$.  Conditional on a $\nu^{(t-1)}$ and $Z_g^{(t-1)}$, we may use
$Z_g^{\mathrm{prior}}$ and ultimately Eq.~(\ref{eq:marg}) in the likelihood
ratio ({\em bottom-right} bubble in Figure
\ref{f:ess}) via $Z_g^{(t)} \rightarrow F_g^{(t)} \rightarrow F_n^{(t)} $
and $Z_g^{(t-1)} \rightarrow F_g^{(t-1)} \rightarrow F_n^{(t-1)}$ to
eventually accept $Z_g^{(t)}$. Finally, conditional on $Z_g^{(t)}$ via
$F_n^{(t)}$, a random-walk proposal such as $\nu' \sim
\mathrm{Unif}[\nu^{(t-1)}/2, 2\nu^{(t-1)}]$ may be accepted as $\nu^{(t)} =
\nu'$ with probability
\begin{equation}
\alpha(\nu^{(t)}, \nu') = 
\frac{p(Y_n \mid F_n^{(t)}, \nu')}{p(Y_n \mid F_n^{(t)}, \nu^{(t-1)})}
\times \frac{\mathrm{Gamma}(\nu'; \alpha_\nu, \beta_
\nu)}{\mathrm{Gamma}(\nu^{(t-1)}; \alpha_\nu, \beta_\nu)} 
\times \frac{\nu^{(t-1)}}{\nu'}, \label{eq:mhnu}
\end{equation}
or otherwise $\nu^{(t)} = \nu^{(t-1)}$.    Posterior sampling for $\theta$ is similar.  Conditional on all other quantities, a proposal
$\theta' \sim
\mathrm{Unif}[\theta^{(t-1)}/2, 2\theta^{(t-1)}]$ may be accepted as $\theta^{(t)} =
\theta'$ with probability
\begin{equation}
\alpha(\theta^{(t)}, \theta') = 
\frac{p(Z_g^{(t)} \mid \theta')}{p(Z_g^{(t)} \mid \theta^{(t-1)})}
\times \frac{\mathrm{Gamma}(\theta'; \alpha_\theta, \beta_\theta)}{\mathrm{Gamma}(\theta^{(t-1)}; \alpha_\theta, \beta_\theta)} 
\times \frac{\theta^{(t-1)}}{\theta'},  \label{eq:mhtheta}
\end{equation}
where $p(Z_g \mid \theta)$ is the pdf of $\mathcal{N}_{n_g}(Z_g ; 0_{n_g}, C_g)$ where $C_g$ is built from $\theta$. The entire Metropolis-within-Gibbs algorithm is encapsulated in Algorithm \ref{alg:gibbs} for convenience.

\medskip
\begin{algorithm}[ht]    
\DontPrintSemicolon
Initialize $Z_g^{(0)}$, $\theta^{(0)}$, $\nu^{(0)}$. \\
\For{$t = 1, \dots, T$}{
  Sample $Z_g^{(t)} \mid Z_g^{(t-1)}, \nu^{(t-1)}, \theta^{(t-1)}$ via ESS (Figure~\ref{f:ess}),
  marg.~lik.~(\ref{eq:marg}) and $(\hat{\mu}, s^2)$ (\ref{eq:summ}). \\
  Sample $\nu^{(t)} \mid Z_g^{(t)}$ via MH (\ref{eq:mhnu}), marg.~lik.~(\ref{eq:marg}) and $(\hat{\mu}, s^2)$ (\ref{eq:summ}). \\
  Sample $\theta^{(t)} \mid Z_g^{(t)}, \theta^{(t-1)}$ via MH (\ref{eq:mhtheta})with MVN pdfs using $C_g$ for each $\theta$.
  }
\caption{MCMC (Metropolis-within-Gibbs) for mono-GP.}
\label{alg:gibbs}
\end{algorithm}

Prediction at $\mathcal{X}_{n'}$ is a simple matter of running back through the MCMC samples, possibly after discarding for burn-in and thinning, to form
\begin{align}
\mathcal{F}_{n'}^{(t)} &= \mbox{\tt monoref}(\mathcal{X}_{n'}, X_g, Z_g^{(t)})
\nonumber \\ 
\mbox{ yielding } \quad 
\mathcal{Y}_{n+1}^{(t)} &\sim 
\mathrm{St}_{n'}( \hat{\mu}^{(t)} + \nu^{(t)}( \mathcal{F}_{n'}^{(t)} - \textstyle\frac{1}{2}), \mathbb{I}_{n'} (1 + \textstyle\frac{1}{n}) s^{2(t)}, n-1), \label{eq:predbayes}
\end{align}
where St is a location-scale Student-$t$ distribution with degrees of freedom
$n-1$.  Moments $\hat{\mu}^{(t)}$ and $s^{2(t)}$ follow Eq.~(\ref{eq:summ})
with $ \hat{\nu}^{(t)}$ and $F_n^{(t)}$ values plugged in.  Although written
implying a multivariate density, a diagonal covariance means it
can be handled element-wise in scalar form.   For most
applications, full posterior predictive samples (\ref{eq:predbayes}) are overkill.
When benchmarking via RMSE or scoring rules, we use moment-based
aggregates:
\begin{align}
\hat{\mu}(\mathcal{X}_{n'}) &= \frac{1}{T} \sum_{t=1}^T \label{eq:predmom}
\hat{\mu}^{(t)} + \nu^{(t)}( \mathcal{F}_{n'}^{(t)} - \textstyle\frac{1}{2}) \\
\Sigma(\mathcal{X}_{n'}) &= 
\mathbb{C}\mathrm{ov} \left[ \hat{\mu}^{(t)} + \nu^{(t)}( \mathcal{F}_{n'}^{(t)} - \textstyle\frac{1}{2}) \right]_{t=1}^T + \mathbb{I}_{n'} \times \frac{1}{T} \sum_{t=1}^T s^{2(t)} \times \frac{(n-1)}{(n-3)}. \nonumber
\end{align}
Notice how these reveal spatial dependence in predictions explicitly through
the covariances on the mean via the law of total variance.  The change in
denominator from $n-1$ to $n-3$ arises from the expression for Student-$t$
variance.  All predictive calculations are linear in $n$, $n'$ and $T$ except
$\mathbb{C}\mathrm{ov}$ in Eq.~(\ref{eq:predmom}) provided posterior samples
following Alg.~\ref{alg:gibbs}.  If the $\mathcal{O}({n'}^2)$ cost of
$\mathbb{C}\mathrm{ov}$ is prohibitive, a linear-cost pointwise variance may
be calculated instead.

\subsubsection*{An illustration}

Recall the visual provided in Figure \ref{f:essviz}.  The true
data-generating mechanism is
$$
Y(x) = \frac{10 \exp(10x - 5)}{1 + \exp(10x - 5)} + \varepsilon_i 
\quad  \mbox{ where } \quad \varepsilon_i \stackrel{\mathrm{iid}}{\sim} \mathcal{N}(0, 1) \quad \mbox{ and } x \in [0,1].
$$
In other words, it is within our assumed model class with $(\mu,\nu, \sigma^2)
= (0,10,1)$, and $f(x) = \mathrm{logit}^{-1}(10x-5)$.  For posterior sampling,
and actually for all MCMC in Sections \ref{sec:mono}--\ref{sec:add}, we take
$T=5000$ total samples, discard the first $B=1000$ as burn-in and thin to take
every $10^\mathrm{th}$, implicitly re-defining $T=400$.  Throughout we use
$X_g \in [0,1]$ with $n_g = 50$.  The right panel of Figure \ref{f:essviz}
shows {\tt monoref} versions of ESS samples along $X_g$.  Note there are fewer
dashed lines in the right panel, indicative of faster ESS acceptance.%However, recall that the visuals in Figure \ref{f:essviz} used $X_g$. 
% Observe that those latent functions ($F_g^{(t)}$) are monotonic by
% construction, and that convergence and ESS acceptance is faster than the 
% ordinary (left panel).

\begin{figure}[ht!]
\centering
%\fbox{
\begin{minipage}{7.5cm}
\includegraphics[scale=0.7, trim=5 15 15 50,clip=TRUE]{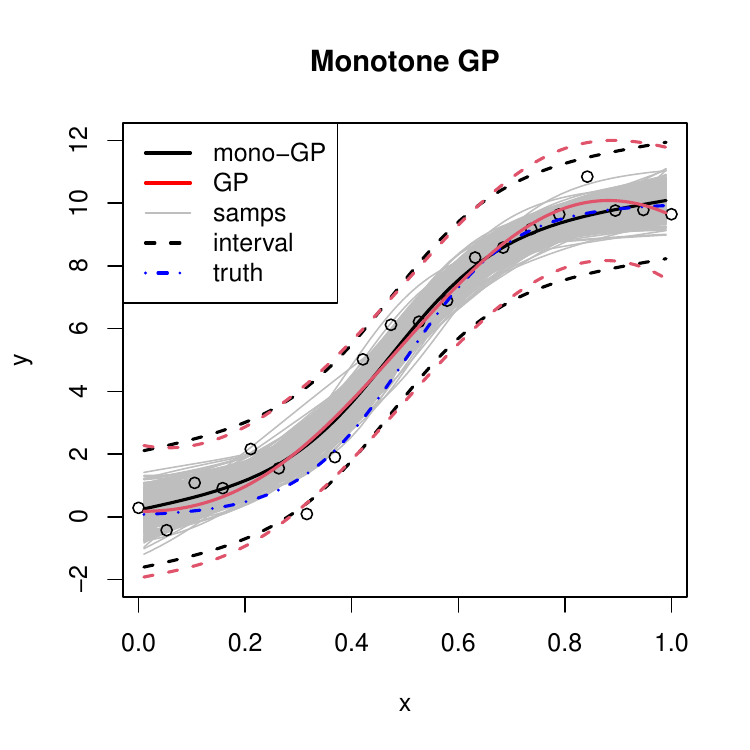}
\end{minipage} \hspace{0.75cm} % } \fbox{
\begin{minipage}{7cm}
\includegraphics[scale=0.65, trim=3 65 0 50,clip=TRUE]{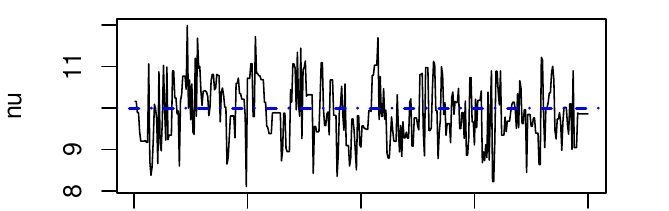}
\includegraphics[scale=0.65, trim=3 10 0 40,clip=TRUE]{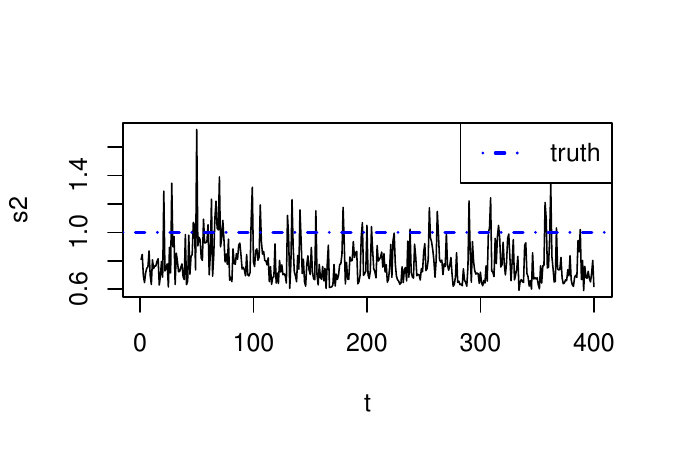}
\end{minipage} % }
\caption{Predictive surfaces (left), with mono-GP posterior samples $F_n^
{(t)}$ as gray lines, and MCMC trace-plots (right) for 1d logistic example.
\label{f:logit1}}
\end{figure}

The left panel of Figure \ref{f:logit1} completes the picture by showing
samples from the (shifted $\mu$ and scaled $\nu$) latent function
$\mathcal{F}_{n'}$ in gray, along with predictive means and quantiles derived
from posterior averages of Eq.~(\ref{eq:predbayes}).  Predictive
$\mathcal{X}_{n'}$ vary by example, and for this $\mathrm{logit}^{-1}$ one we
used a regular $n'=100$-sized grid.  For comparison, a regular GP -- with
analytically marginalized latents via {\tt laGP} \citep{laGP} -- is shown in
red.  While the GP is highly accurate, it is not monotonic.  The trace plots
in the right panel(s) of Figure \ref{f:logit1} show that MCMC mixing is good
for $\nu^{(t)}$ and for (the derived quantity) $s^{2(t)}$, and they also hover
around their true values. Trace plots for $\mu^{(t)}$ are similar, but not
shown.  Traces $\theta^{(t)}$ also indicate good mixing (not shown), but there
is no true value to compare to directly. Latent $\mathcal{F}_{n'}^{(t)}$
(gray/left panel), which come indirectly from those $\theta^{(t)}$, capture the
truth well.

\section{Additive framework for multiple inputs}
\label{sec:add}

Extending monotonicity to more inputs is nuanced since there is no natural
ordering on $\mathbb{R}^p$ for $p \geq 2$.  Here we consider
the simplest possible extension of coordinate-wise monotonicity.
%, which is the same as monotonic along straight lines (with increasing coordinates).

\subsection{Bayesian modeling and inference}

We capture this coordinate-wise notion in a statistical model as
follows.  Let $X_n$ be $n \times p$ with rows $x_i^\top = (x_{i1,}, \dots,
x_{ip})^\top$, for $i=1, \dots n$.  Then presume $y_i = \mu + \sum_{j=1}^n
\nu_j f_j(x_{ij}) + \varepsilon_i$, with $\varepsilon_i
\stackrel{\mathrm{iid}}{\sim} \mathcal{N}(0, \sigma^2)$ where $f_j(x_{ij})
\leq f_j(x_{kj})$ for $x_{ij} \leq x_{kj}$, and $v_j > 0$ for all $j =1, \dots, p$. In
other words, all $f_j$ are 1d monotonic functions, and their linear
combination involves only positive coefficients $\nu_j$.  Observe that the
$p=1$ case of Section \ref{sec:mono} is nested within this setup.

This setup also resembles \citet{lopez2022high}, but our fully hierarchical
model and inferential apparatus are quite distinct.  We prescribe $p$
independent but otherwise identical copies of everything in
Eq.~(\ref{eq:model}) except the likelihood (which is unchanged).  That is, let
$F_n$ be $n \times p$ like $X_n$ where the $j^\mathrm{th}$ column of $F_n$
corresponds to the $j^\mathrm{th}$ {\em a priori} independent process on the
$j^\mathrm{th}$ column $X_n^{\cdot j}$ of $X_n$.  Then let $\nu = (\nu_1,\dots,
\nu_p)^\top$ denote a $1 \times p$ row-vector of {\em a priori} independent
scales.  Using these quantities, interpret the likelihood in Eq~(\ref{eq:model})
as a vector-matrix product in $\nu (F_n - \textstyle \frac{1}{2})$, where
$\frac{1}{2}$ subtracts off every element of $F_n$.

Duplicating (actually $p$-licating) the other lines of
Eq.~(\ref{eq:model}) requires some care.  We use an identical 1d reference
grid $X_g$ for each coordinate, $j=1,\dots,p$, but with distinct lengthscales
$\theta=(\theta_1,
\dots, \theta_p$), in keeping with an automatic relevance determination
\citep[ARD;][]{williams2006gaussian}-like setup.  So the full set of unknowns
is $\mu,
\sigma^2, \nu, \theta$ as before, but $\nu$ is now vectorized. This makes
inference a straightforward extension of Section \ref{sec:monopost}. MCMC
basically follows Alg.~\ref{alg:gibbs} in $p$-licate via the following summary
statistics extending Eq.~(\ref{eq:summ}).
\begin{align}
\hat{\mu} &= \frac{1}{n} \sum_{i=1}^n y_i - \sum_{j=1}^p\nu_j(f_{ij} - \textstyle \frac{1}{2}) 
&& \mbox{ and } &
s^2 &= \frac{1}{n-p} \sum_{i=1}^n (y_i - \hat{\mu} - \sum_{j=1}^p \nu_j(f_{ij} - \textstyle\frac{1}{2}))^2 \label{eq:addsumm}
\end{align}
Above, $\nu (F_n - \frac{1}{2})$ is written out explicitly in sum-form for
clarity.  Similarly, predictions via
Eq.~(\ref{eq:predbayes}--\ref{eq:predmom}) may be applied identically with
$n-p$ for $n-1$, including $n-p-2$ for $n-3$.  Again, this setup nests $p=1$
as a special case.  
% The simplicity of this extension of the 1d setup is both
% aesthetically pleasing and makes for a simple implementation.

\subsubsection*{An illustration}

Consider a $p=2$ upgrade to our running logistic example from earlier
sections.  We shall benchmark in higher dimension momentarily in Section
\ref{sec:monobench}. Take $f_1(x) = \mathrm{logit}^{-1}(10x - 7)$ and
$f_2(x)=
\mathrm{logit}^{-1}(10x-3)$ with $\nu = (10, 5)$, differentially shifting and
scaling the 1d version. Complete the setup with $(\mu, \sigma^2)=(0, 0.1)$.
Training $X_n$ are taken from a Latin hypercube sample
\citep[LHS;][]{mckay2000comparison} of size $n=100$ in 2d.
\begin{figure}[ht!]
\centering
\includegraphics[scale=0.55, trim=0 65 15 5,clip=TRUE]{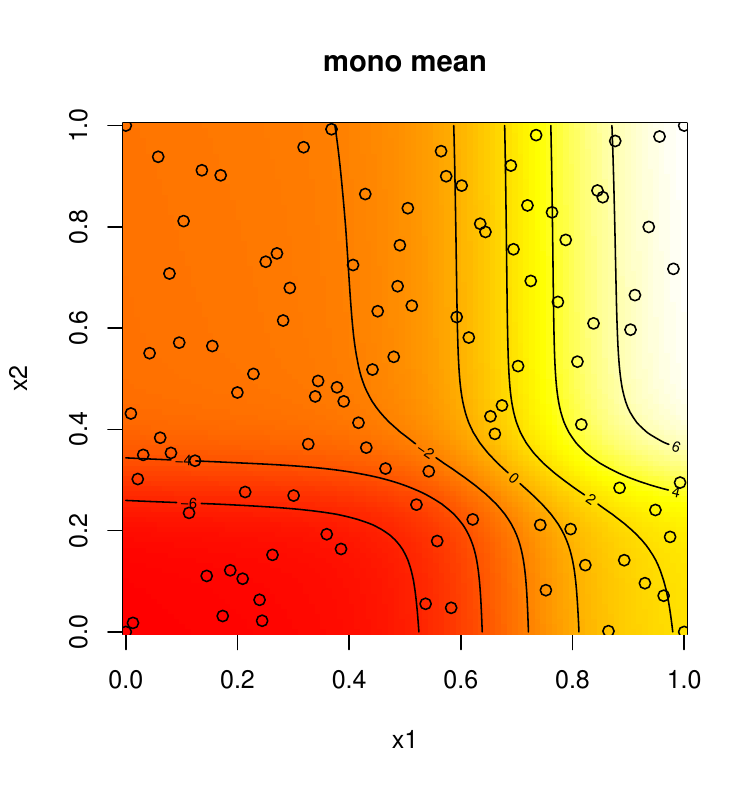}
\includegraphics[scale=0.55, trim=50 65 0 5,clip=TRUE]{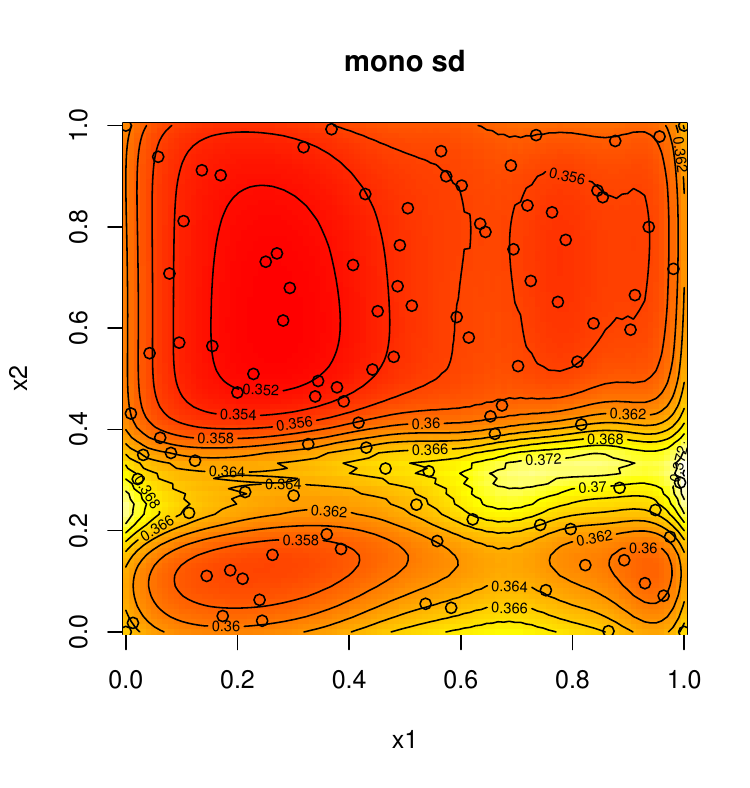}
\includegraphics[scale=0.55, trim=0 15 15 5]{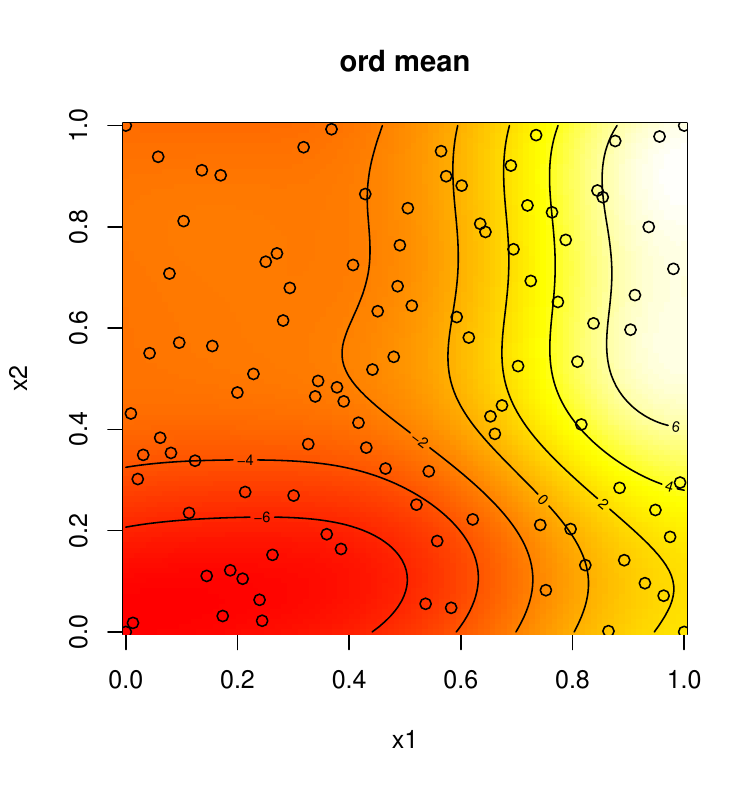}
\includegraphics[scale=0.55, trim=50 15 0 5,clip=TRUE]{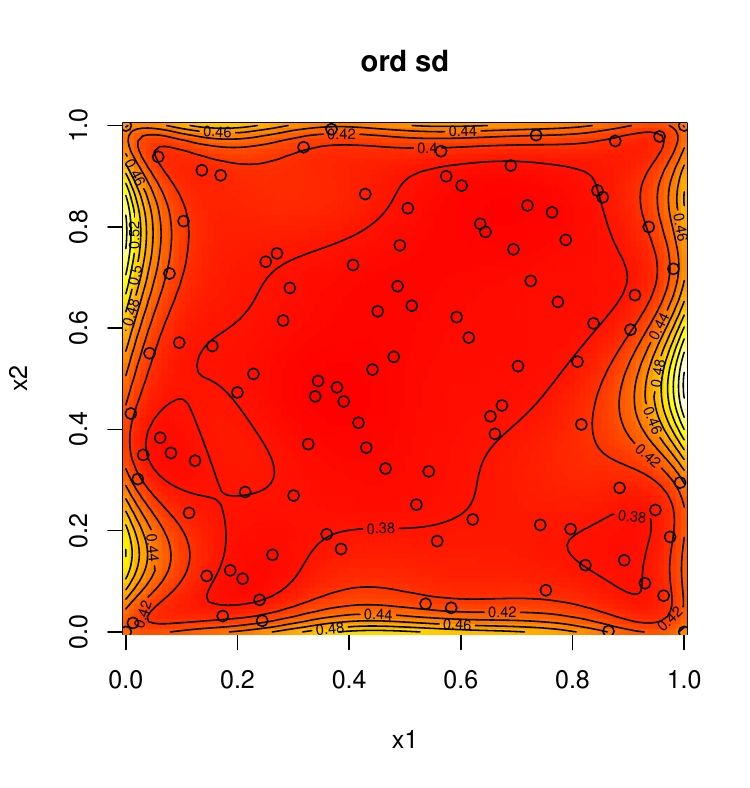}
\caption{2d logistic fits from a monotonic additive GP (mono-GP,
{\em top row}) to an ordinary GP (via {\tt laGP}, {\em bottom row}) on mean
({\em left}) and SD ({\em right}).  Red is low and white is high, with novel
color palettes in each panel.
 \label{f:logit2surfs}}
 \end{figure}
We then make predictions on a $100 \times 100$ grid $\mathcal{X}_{n' =
10\mathrm{K}}$ to support visuals like those in Figure \ref{f:logit2surfs}
which offer a comparison between our monotonic additive GP (still mono-GP, {\em
top row}) and an ordinary GP (with ARD, {\em bottom row}).  Focus first on the
{\em left} column, showing predictive means.  Despite superficial
similarities, notice how contour lines for the ordinary GP wiggle: telltale
departure from monotonicity.  Now look at the the standard deviations on the
{\em right}. The ordinary GP exhibits typical behavior with increasing
uncertainty in distance to nearby training data (open circles). Mono-GP
exhibits a more nuanced spatial UQ, seeming to be aligned with regions where
the response is changing most quickly.

\begin{figure}[ht!]
\centering
\includegraphics[scale=0.65, trim=0 15 15 55,clip=TRUE]{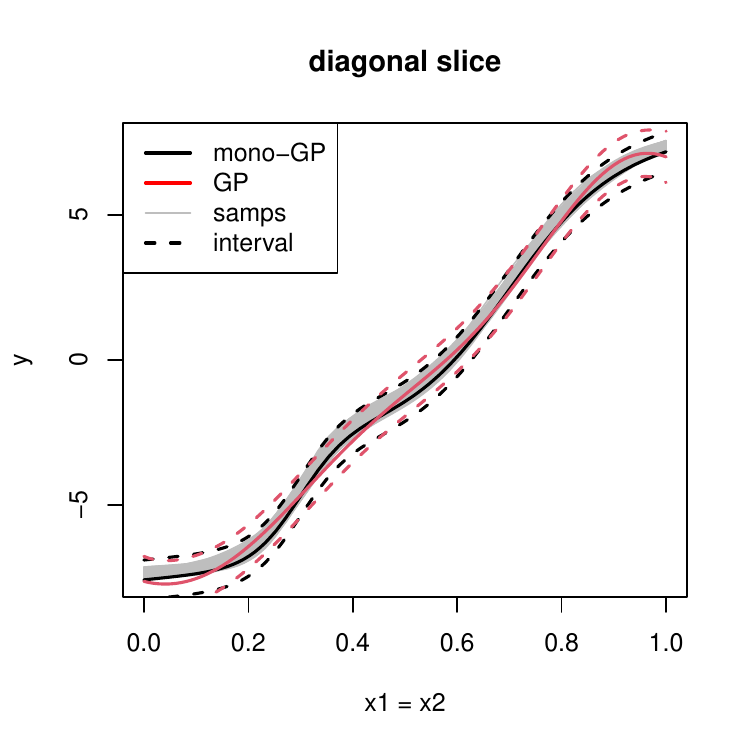}
\includegraphics[scale=0.65, trim=20 15 10 55,clip=TRUE]{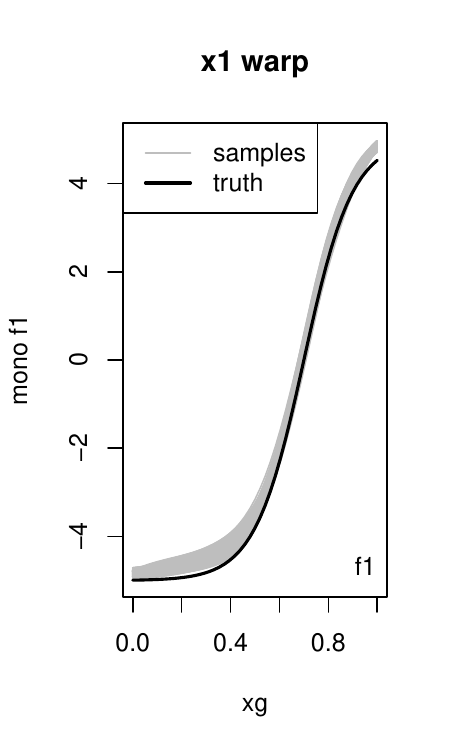}
\includegraphics[scale=0.65, trim=50 15 0 55,clip=TRUE]{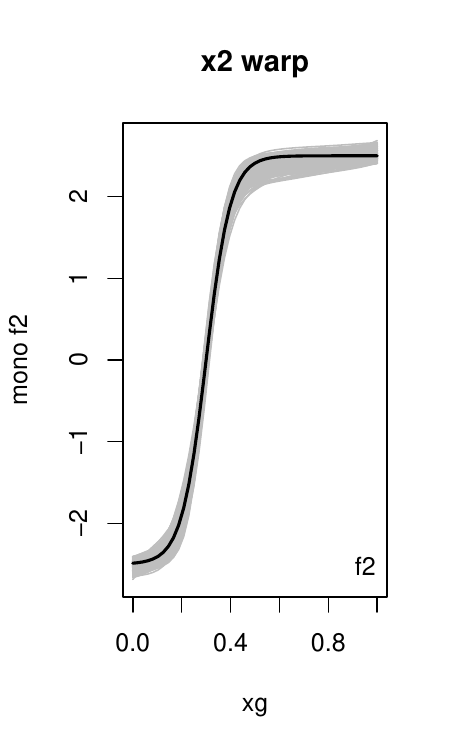}
\caption{Diagonal slice of 2d logistic predictive surface ({\em left}) and
mono-GP latent functions in each input compared to the truth ({\em middle} and
{\em right}). \label{f:logit2break}}
\end{figure}

Figure \ref{f:logit2break} shows three 1d views, first comparing mono-GP to
the ordinary GP along the $x_1 = x_2$ slice, and then coordinate-wise for
columns of $F_n$ over $X_g$.  Observe on the {\em left} a view similar to the
left panel of Figure \ref{f:logit1}.  As in that case, the surfaces are
similar but the one from the ordinary GP is not monotonic, particularly at the
two corners of the input space.  The {\em middle} and {\em right} panels show
samples of columns of $F_n$ relative to the truth: our latent functions are
generally capturing the true data-generating mechanism.

\subsection{Implementation and benchmarking}
\label{sec:monobench}

Here we upgrade our earlier, largely visual illustrations into a series of
benchmarking exercises.  Metrics include RMSE to the truth, and continuous
ranked probability score \citep[CRPS;][]{gneiting2007strictly} to noisy
testing realizations.  Our implementation of mono-GP requires about 200 lines
of pure {\sf R} code, using add-on libraries only for MVN deviates via {\tt
mvtnorm} on CRAN \citep{mvtnorm} and distance calculation via {\tt laGP}.  Our
monotonic competitor, lineq-GP, is facilitated by {\sf R} code provided by
\citet{lopez2022high}, via {\tt lineqGPR} \citep{lineqGPR}, without
modification.  As earlier, we use {\tt laGP}, which is predominantly
in {\sf C}, for an ordinary GP. The 1d logistic example from earlier is
relegated to Appendix \ref{app:logistic} because it is uninteresting. Appendix
\ref{app:timing} provides timing comparisons varying $n$ and $n'$, however
pitting {\sf C} against {\sf R} is like apples to oranges.

\subsubsection*{2d logistic}

We begin  with the 2d logistic variation introduced above in Section
\ref{sec:add}.  The experimental setup here is identical to that one, where
each new Monte Carlo (MC) instance uses a novel LHS of size $n=100$ paired
with novel error deviates on the response.  We use 30 MC repetitions throughout.
This 2d example uses the fixed $n' = 10\mathrm{K}$ testing grid described above,
whereas our other higher dimensional examples use LHSs.

\begin{figure}[ht!]
\centering
\includegraphics[scale=0.65, trim=0 5 0 50,clip=TRUE]{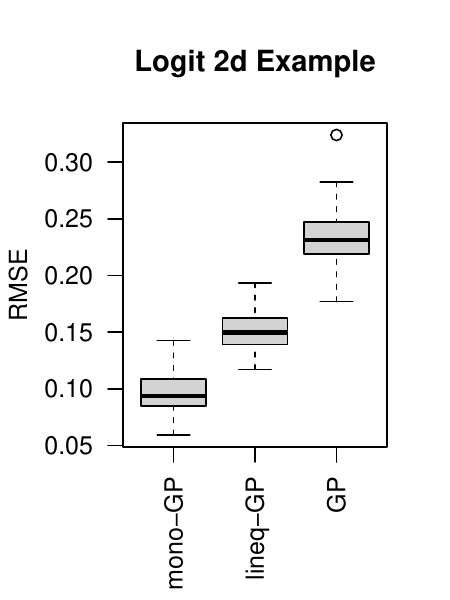}
\includegraphics[scale=0.65, trim=0 5 10 50,clip=TRUE]{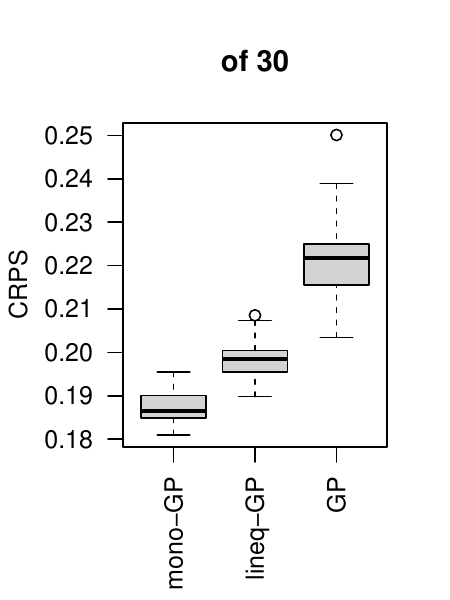}
\vspace{-0.1cm}
\caption{Metrics on the 2d logit MC experiment.  Smaller is better for all.
\label{f:logit2bench}}
\end{figure}

Figure \ref{f:logit2bench} collects the results.  Observe that the monotonic
methods outperform an ordinary GP (lower RMSE/CRPS), and that our mono-GP
edges out the lineq-GP from \citeauthor{lineqGPR}.  It is worth nothing that {\tt
lineqGPR} was not created for bakeoffs of this kind. The method targets a
wider class of constrained problems and emphasizes active learning
applications. 

% Likewise, our mono-GP method was actually created for the DGP application in
% Section \ref{sec:mwDGP}.

\subsubsection*{5d Lopez--Lopera}

\begin{figure}[ht!]
\centering
\includegraphics[scale=0.65, trim=0 5 0 50,clip=TRUE]{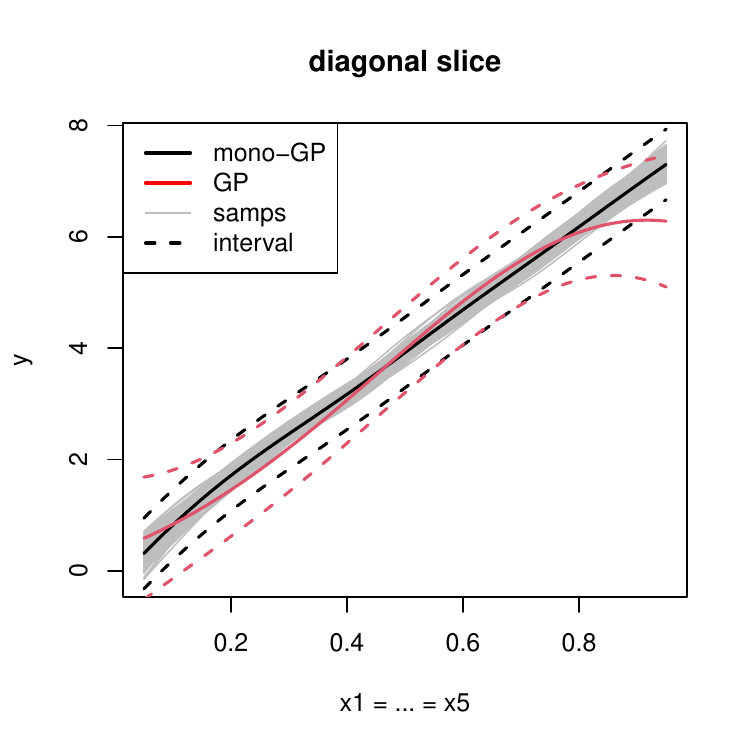}
\hspace{0.5cm}
\includegraphics[scale=0.65, trim=20 5 25 50,clip=TRUE]{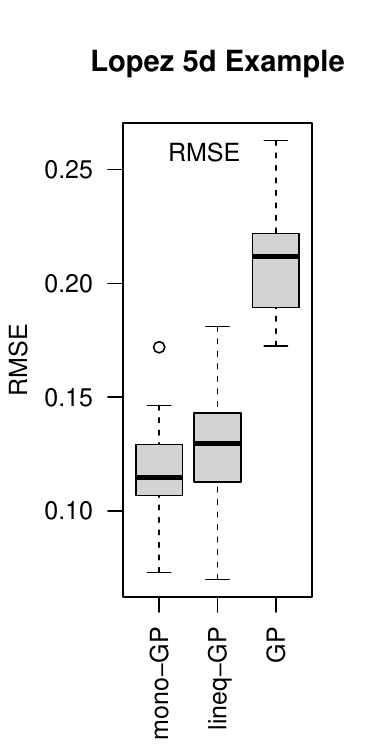}
\includegraphics[scale=0.65, trim=15 5 5 50,clip=TRUE]{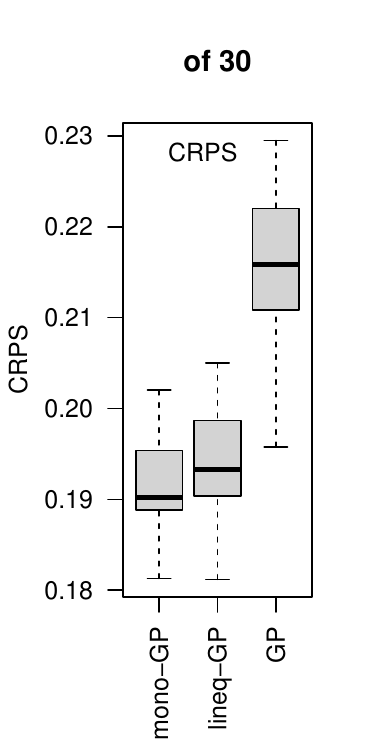}
\caption{Diagonal visual ({\em left}) and metrics on the 5d Lopez--Lopera MC experiment.
\label{f:lopez5bench}}
\end{figure}

Our remaining examples come from \citet{lopez2022high}~and use $p=5$ and
$p=10$, respectively.  Both can be seen as inherently logistic in nature,
though with slightly different emphasis on coordinates.  The first one follows
$$
y(x) = \mathrm{atan}(5 x_1) + \mathrm{atan}(2 x_2) + x_3 + 2 x_4^2
+ \frac{2}{1 + \exp(-10 (x_5 - \textstyle\frac{1}{2}))} + \varepsilon, \quad
\varepsilon \stackrel{\mathrm{iid}}{\sim} \mathcal{N}(0,0.1^2).
$$
Like the 2d logistic above, this can be re-expressed in our additive model
class.   Figure \ref{f:lopez5bench} summarizes the
outcome of a MC experiment based on novel $n=100$ training and $n'=1000$
testing LHSs.
Observe that mono-GP and lineq-GP are similar on RMSE and CRPS, yet in only
one of the thirty reps was mono-GP inferior to lineq-GP on either metric.
Consequently, a one-sided paired Wilcoxon test rejects similarity with $p=1.3
\times 10^{-8}$.  The left panel shows predictive surfaces along the diagonal,
confirming that the ordinary GP is unable to learn that the response surface
is non-decreasing.

\begin{figure}[ht!]
\centering
\includegraphics[scale=0.65, trim=30 15 30 50,clip=TRUE]{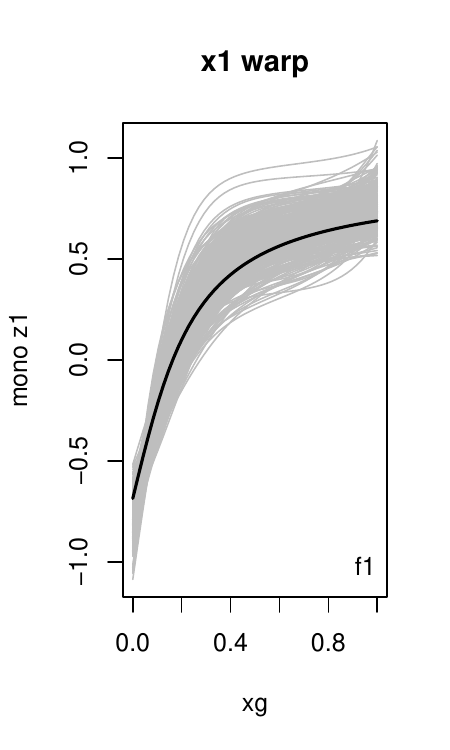}
\includegraphics[scale=0.65, trim=55 15 30 50,clip=TRUE]{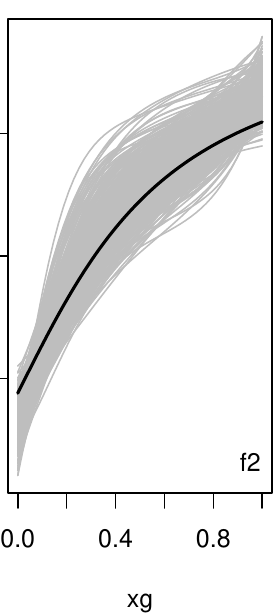}
\includegraphics[scale=0.65, trim=55 15 30 50,clip=TRUE]{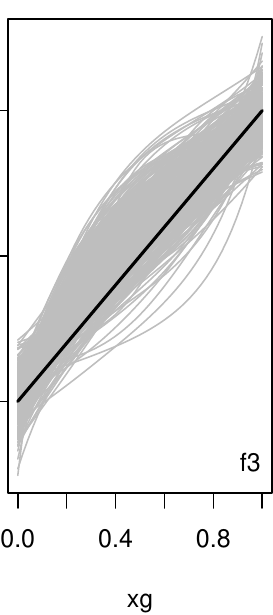}
\includegraphics[scale=0.65, trim=55 15 30 50,clip=TRUE]{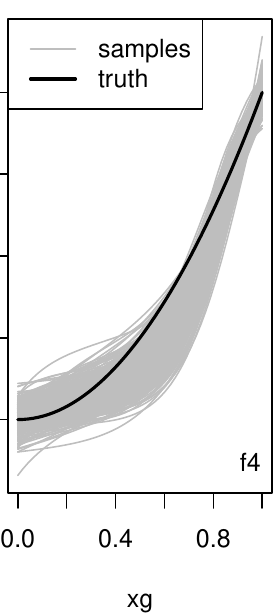}
\includegraphics[scale=0.65, trim=55 15 30 50,clip=TRUE]{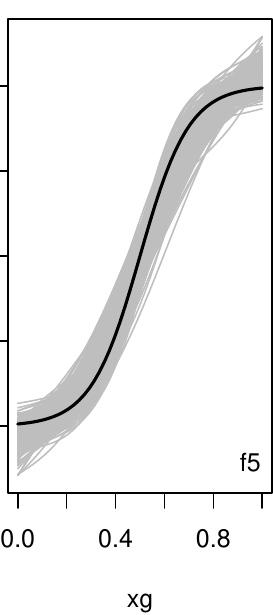}
\caption{Comparison of mono-GP latent functions for 5d Lopez--Lopera example to the truth. \label{f:lopez5dlatent}}
\end{figure}

Figure \ref{f:lopez5dlatent} shows the latent functions learned for each input
coordinate compared to the truth.  It is nice to have this interpretive aspect
to our mono-GP implementation.  We could couple that with the amplitudes
provided by $\nu^{(t)}$ [see Appendix \ref{app:sens}], but we prefer here to
show each on the same $y$-axis. With ordinary GPs, it can be difficult to
tease out the relative contributions of each input, say via an {\em ex post}
sensitivity analysis \citep[e.g.,][Chapter
9.1]{saltelli2000sensitivity,marrel2009calculations,oakley2004probabilistic,gramacy2020surrogates}.
With mono-GP a coordinate-wise breakdown is implicit in its construction.

\subsubsection*{10d Lopez--Lopera arctan}

\begin{figure}[ht!]
\centering
\includegraphics[scale=0.65, trim=0 5 0 50,clip=TRUE]{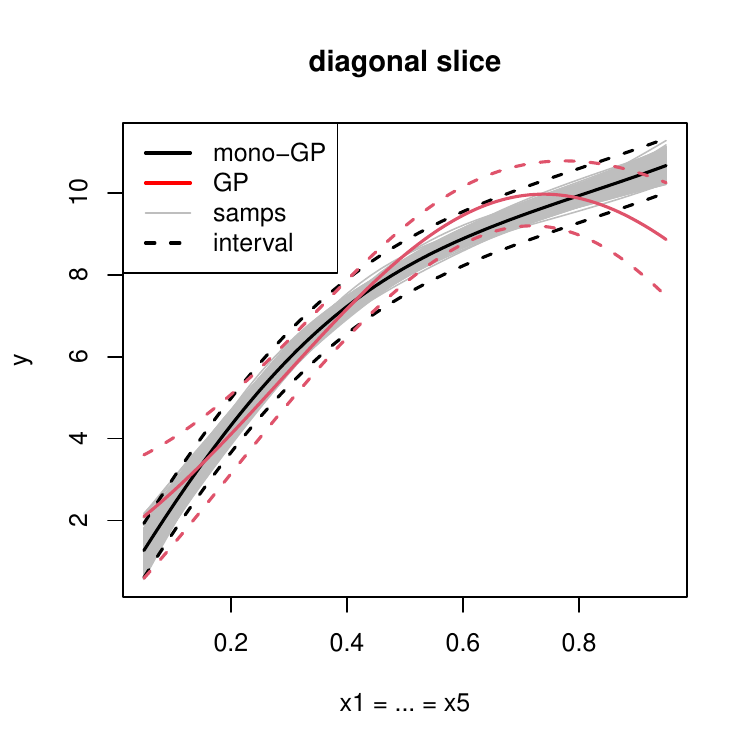}
\hspace{0.5cm}
\includegraphics[scale=0.65, trim=20 5 25 50,clip=TRUE]{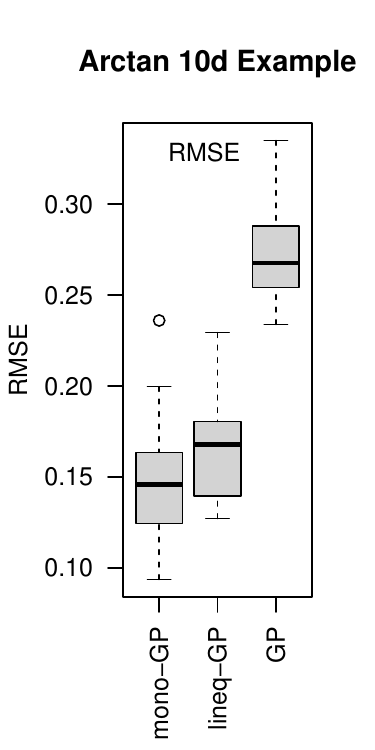}
\includegraphics[scale=0.65, trim=15 5 5 50,clip=TRUE]{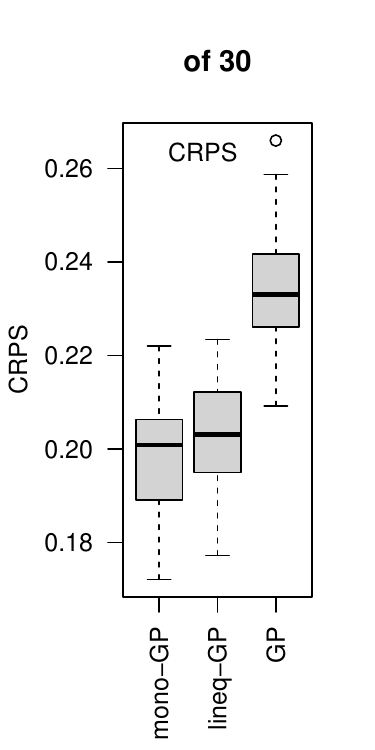}
\caption{Diagonal visual ({\em left}) and metrics on the 10d Lopez--Lopera MC experiment.
\label{f:arctan10bench}}
\end{figure}

Our next \citeauthor{lopez2022high} example is  originally from
\citet{bachoc2022sequential}.
$$
y(x) = \sum_{j=1}^p \mathrm{atan}\left(5 \left[1 - \frac{1}{p+1} \right] x_j \right) + \varepsilon, \quad
\varepsilon \stackrel{\mathrm{iid}}{\sim} \mathcal{N}(0,0.1^2)
$$
Here we consider the $p=10$ case.  Other things from previous experiments are
unchanged, and Figure \ref{f:arctan10bench} summarizes the results of our MC
benchmarking exercise.
Again, it is obvious that an ordinary GP is inferior (boxplots on {\em right})
because it is unable to leverage that the response surface is monotonic ({\em
left}).  Mono-GP and lineq-GP are similar, but again mono-GP is only inferior
twice out of thirty times by either metric leading to $p= 9.3 \times 10^{-9}$,
rejecting parity.  A coordinate-wise sensitivity analysis via columns of $F_n$
is provided in Appendix \ref{app:sens}.

\section{Monotonic warpings for deep GPs}
\label{sec:mwDGP}

Deep Gaussian processes (DGPs) originate in spatial statistics
\citep[e.g.,][]{sampson1992nonparametric,schmidt2003bayesian} for
nonstationary modeling
\citep[e.g.,][]{paciorek2003nonstationary,sauer2023non}, and have been
recently re-popularized in ML
\citep{salimbeni2017doubly,bui2016deep,cutajar2017random,laradji2019efficient}
by analogy to deep neural networks (DNNs).  Here we focus on
\citet{sauer2023active} whose fully Bayesian DGP provides the UQ
necessary for many surrogate modeling applications of interest to us.
\citeauthor{sauer2023active}'s ESS-based inferential framework laid the
groundwork for our monotonic setup in Section \ref{sec:mono1}. We provide a
brief DGP review here, with the minimum detail required to explain our
contribution.  As a disclaimer, note that we are pivoting back to ordinary
(not monotonic) regression, with mono-GP featuring as an important subroutine.

All of the modeling ``action'' for a conventional GP, generically $Y \sim
\mathcal{N}(0,
\Sigma(X))$, resides in $\Sigma(X)$.  When the kernel depends on displacement
only, i.e., $\Sigma^{ij} = \Sigma(x_i - x_j)$, the resulting process is said
to be {\em stationary}.  This means that the positions of inputs in $X$ don't
matter, only the gaps between them.  Consequently, when making predictions at
new inputs $\mathcal{X}_{n'}$, only the displacements between training and
testing sets matter, not their raw locations.  This, in turn, means that you
can't have different ``regimes'' in the input space where input--output
dynamics exhibit diverse behavior.  Stationarity is a strong, simplifying
assumption, that can be prohibitive in many modeling contexts. For example,
fluid dynamics simulations can exhibit both equilibrium and turbulent behavior
depending on their configuration.  A stationary GP could not accommodate both.
We consider an example later that involves modeling a rocket booster
re-entering the atmosphere \citep{pamadi2004aerodynamic} where low-speed lift
dynamics differ from high-speed ones.

A DGP uses two or more stationary GPs in a chain.  One (or more) warps the
inputs into a regime where a stationary relationship is plausible, which is
then fed into another GP for the response, e.g.,  $W_n \sim \mathcal{N}(0,
\Sigma(X_n))$ feeding into $Y_n \sim \mathcal{N}(0, \Sigma(W_n))$.  When there are 
multiple inputs, it is common to match latent and input dimension, specifying $X
\rightarrow W^{\cdot j}$, for $j=1, \dots, p$ with $W^{\cdot j}
\stackrel{\mathrm{iid}}{\sim}
\mathcal{N}(0, \Sigma(X))$.  In other words, the $j^\mathrm{th}$ coordinate of
$W$ is determined by (all of) $X$. Here we restrict our attention to such
``two-layer'' DGPs.   The difficult thing about DGP inference is integrating
out the $n$-dimensional warping variables $W^{\cdot j}$. Twenty years ago
\citet{schmidt2003bayesian} tried Metropolis-based methods, but to say it was
cumbersome is an understatement. Recent ML revivals focus on variational
inference methods with inducing points approximations, but these undercut on
both fidelity and UQ.  For computer simulation experiments,  ESS seems ideal.
MVN priors ($W^{\cdot j} \mid X$) paired with MVN likelihood ($Y
\mid W$) means we can follow Figure \ref{f:ess} in $p$-licate.

\subsection{Component swap}
\label{sec:swap}

Such two-layer DGPs are still supremely flexible.  Every coordinate of $X$
informs each $W^{\cdot j}$, and the relationship between them can be pretty
much any (stationary) function.  Here, we argue that for many computer
simulation experiments some additional regularization is advantageous.  We
propose warping each $X^{\cdot j}$ to $W^{\cdot j}$ via mono-GP rather than a
full GP. In other words, apply the $p$ copies of Section \ref{sec:mono1} for a
coordinate-wise injective warping.  Since only relative distances matter
between (even warped) inputs, we fix $(\mu, \nu, \sigma^2) = (0, 1, 0)$ and we
don't do $- \frac{1}{2}$ so that $W \in [0,1]$.  Since each $W^{\cdot j}$ is
equivalently scaled, the outer GP ($Y\mid W$) requires separable lengthscales,
an upgrade from \citeauthor{sauer2023active}'s original implementation.  
We call this the ``mw-DGP'' where ``mw'' is for ``monowarp''.

\subsubsection*{An illustration}

Consider the ``cross-in-tray'' function, which can be found in the Virtual
Library of Simulation Experiments \citep[VLSE;][]{surjanovic2013virtual}.  We
follow the typical setup
$$
f(x) = - 0.0001 \left(\left| \sin(x_1) \sin(x_2) \exp\left\{\left| 
100 - \frac{\sqrt{x_1^2 + x_2^2}}{\pi}
\right| \right\} \right| +1 \right)^{0.1} \quad \mbox{ with } \quad
x\in[-2,2]^2,
$$
but coded to the unit cube.  A nonstationary GP is essential because there
are both steep (cross) and flat (tray) regions.  We train on an LHS of size $n
= 40$ and predict on a dense 2d grid  of size $n' =900$. Throughout our
DGP-based experiments we use {\tt deepgp}-package defaults (i.e., 10K MCMC
samples), discarding 1K as burn-in and thinning by 10 for $T=900$ total.
Basically we use
\verb!fit_two_layer! with and without a new option: {\tt monowarp = TRUE}.

\begin{figure}[ht!]
\centering
\includegraphics[scale=0.45,trim=0 0 25 55,clip=TRUE]{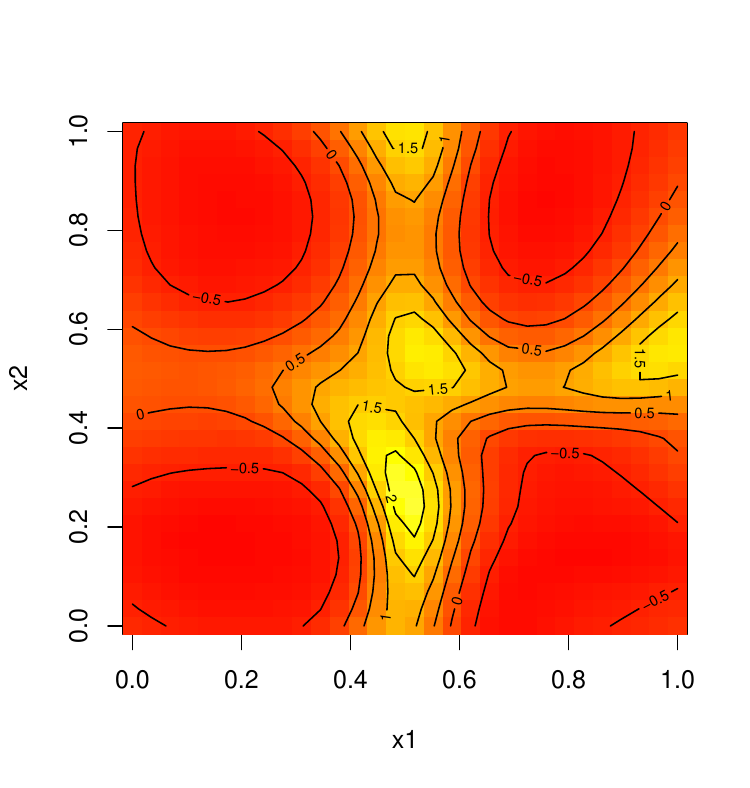}
\includegraphics[scale=0.58,trim=50 55 45 55,clip=TRUE]{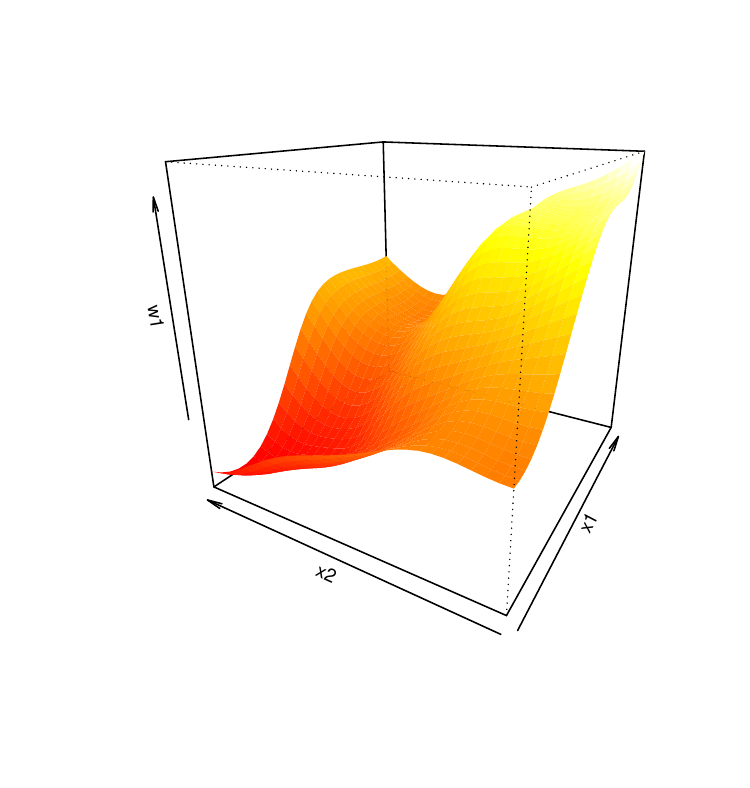}
\includegraphics[scale=0.58,trim=50 55 45 55,clip=TRUE]{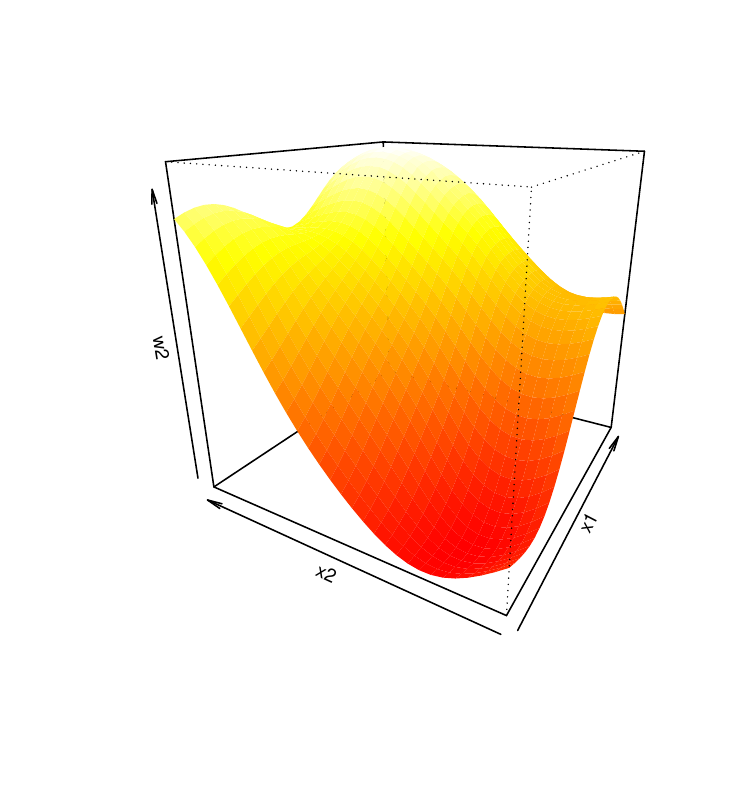}
\includegraphics[scale=0.45,trim=8 15 25 55,clip=TRUE]{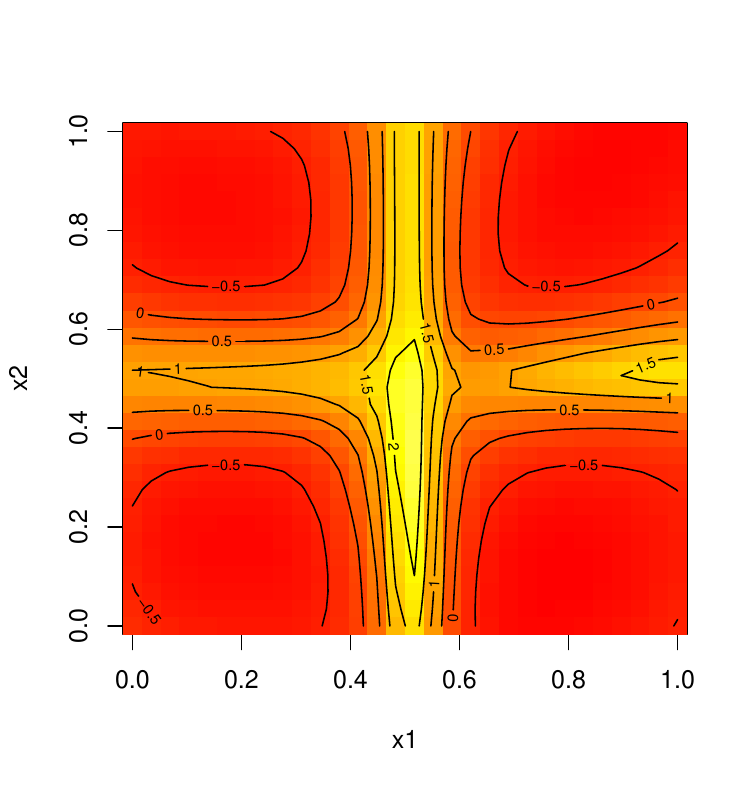}
\includegraphics[scale=0.45,trim=0 15 25 55,clip=TRUE]{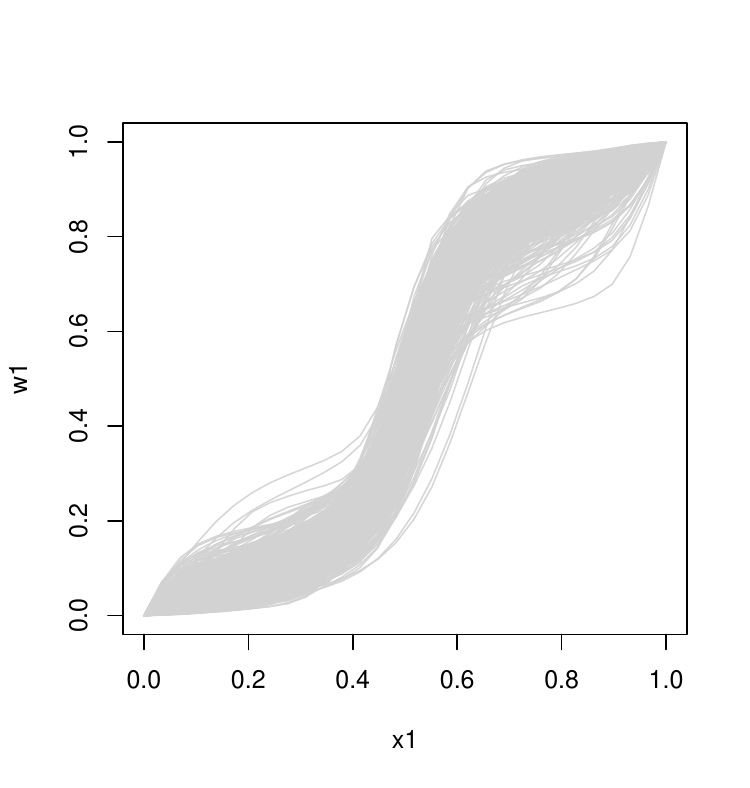}
\includegraphics[scale=0.45,trim=0 15 17 55,clip=TRUE]{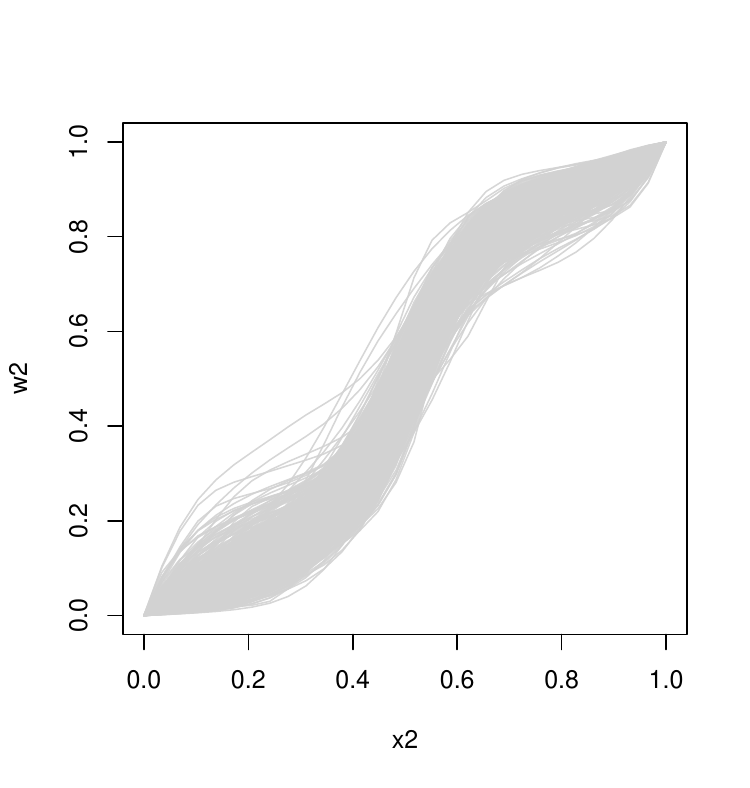}
\caption{DGP (top row) and mw-DGP (bottom row) on cross-in-tray.
Columns provide predictive mean (left) and warpings $W^{\cdot 1}$ and
$W^{\cdot 2}$ (middle and right, respectively). \label{f:cross}}
\end{figure}

Figure \ref{f:cross} summarizes our fits visually, with an ordinary DGP in the
top row of panels and mw-DGP on the bottom.  The left column shows predictive
means.  The true surface makes an axis-aligned cross, so the mw-DGP is doing a
better job of capturing that.  We shall provide full benchmarking results
momentarily.  Observe in the {\em top-right} panels that the warpings from DGP
are not monotonic.   Here we are showing just one sample of $W^{(t)}$ since
these are in 2d.  Observe that both samples have multiple regions where they
fold back on one another.  On the {\em bottom} we show all posterior samples
of coordinate-wise monotonic warpings.  They indicate that the middle of the
input space (the cross) is stretched relative to the edges (tray), which is
supported by the views shown on the left.

\subsection{Benchmarking}
\label{sec:gdpbench}

Here we provide out-of-sample metrics for three examples (with a fourth in
Appendix \ref{app:michael}), all summarized in Figure \ref{f:dgpmets},
beginning with cross-in-tray from Section \ref{sec:swap} in the {\em left/first}
column.  Since we have already introduced that example, we do not provide a
separate heading for it here.  We simply remark that the setup is identical to
that earlier description, except that we repeated it in a MC fashion thirty
times, as with the other examples.

\begin{figure}[ht!]
\centering
\includegraphics[scale=0.75,trim=0 65 20 20,clip=TRUE]{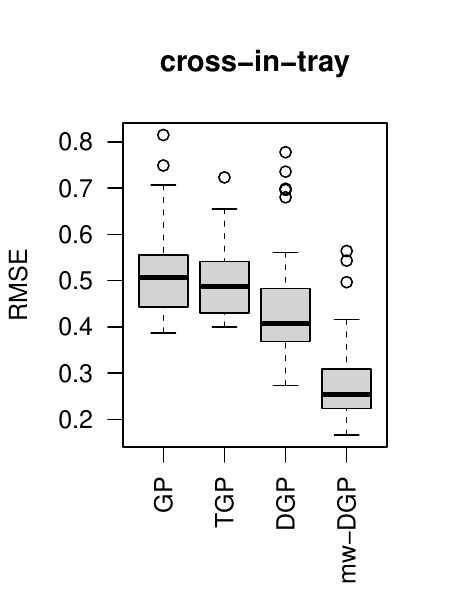}
\includegraphics[scale=0.75,trim=13 65 20 20,clip=TRUE]{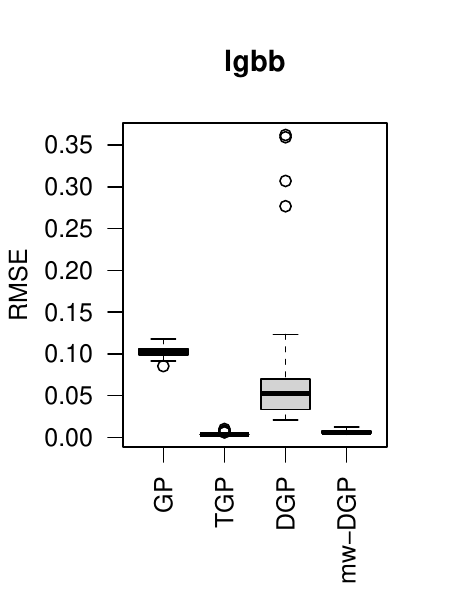}
\includegraphics[scale=0.75,trim=13 65 20 20,clip=TRUE]{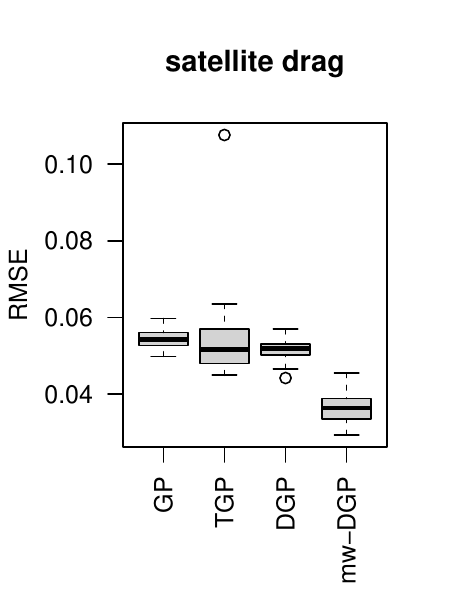}
\includegraphics[scale=0.75,trim=0 5 20 55,clip=TRUE]{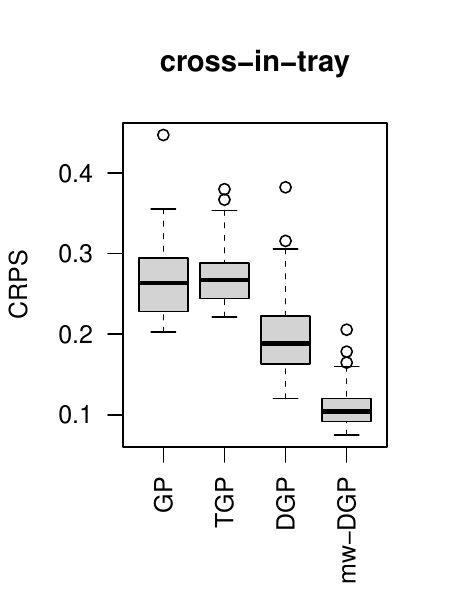}
\includegraphics[scale=0.75,trim=13 5 20 55,clip=TRUE]{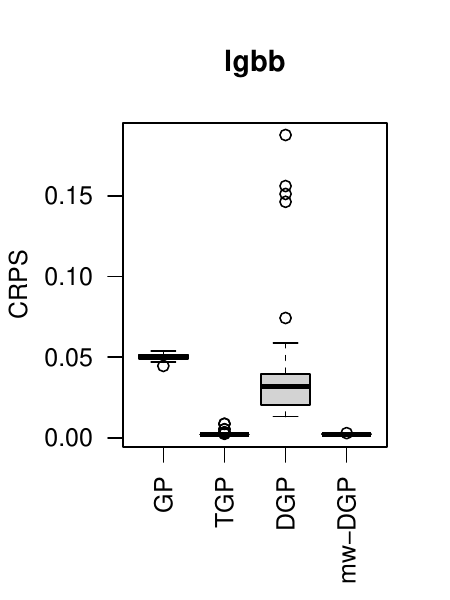}
\includegraphics[scale=0.75,trim=13 5 20 55,clip=TRUE]{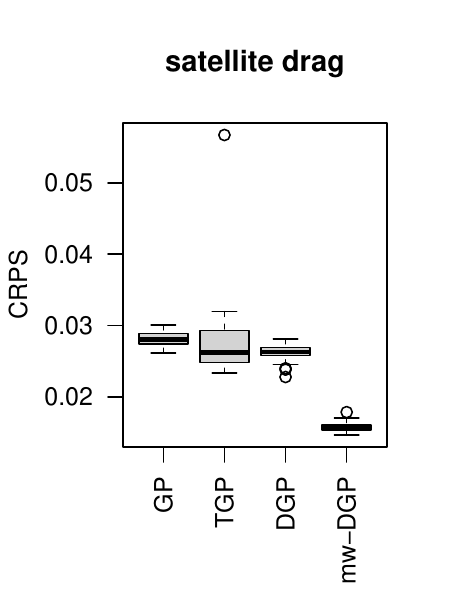}
\caption{Nonstationary benchmark out-of-sample metrics.  Smaller is better
for all.
\label{f:dgpmets}}
\end{figure}

In addition to an ordinary DGP, we include a stationary GP via the same MCMC
apparatus as DGP, using \verb!fit_one_layer! in {\tt deepgp}, and a treed
Gaussian process (TGP) via the {\tt tgp} package on CRAN \citep{tgp1,tgp2}
using {\tt btgpllm} with defaults. The {\em first/left column} of Figure
\ref{f:dgpmets} shows that mw-DGP convincingly outperforms the other
competitors on cross-in-tray. Although mw-DGP utilizes the thriftier
$n_g$-sized reference process for $W$, an ordinary $n$-sized GP is still used
on the outer layer.  Despite being $\mathcal{O}(n^3)$ like the other methods,
many fewer such operations mean a faster execution.  Appendix \ref{app:timing}
provides detailed timings including Vecchia upgrades
\citep{sauer2023vecchia}.

\subsubsection*{Rocket booster}

The second column of Figure \ref{f:dgpmets} is for the Langley Glide-Back
Booster \citep[LGBB;][]{pamadi2004aerodynamic}.  The LGBB simulator models
aeronautically relevant forces as a rocket booster re-enters Earth's
atmosphere as a function of its speed (mach number), angle of attack, and
side slip angle. Here we study the lift force as predicted by those $p=3$
input variables. \citet{gramacy2008bayesian}'s TGP was originally developed
for the LGBB because it exhibits nonstationary dynamics which manifest as
coordinate-wise regime changes in speed and angle of attack. Access to novel
simulations for LGBB is no longer possible (previously, runs could only be
obtained on NASA's Columbia supercomputer), so here we use a corpus of 780
training runs provided by \citet{gramacy2009adaptive} and subsamples of a
densely interpolated set of testing $>37$K outputs provided by a TGP fit from
that same paper.  For more detail see \citet[][Chapter
2]{gramacy2020surrogates}.  Here we use all simulation outputs as training
data $(n = 780)$ and a random subsample of TGP predictions for testing $(n' =
1000)$.

Results are provided in the {\em middle/second column} of panels in Figure
\ref{f:dgpmets}. TGP is a gold standard in this context, having chosen the
training data via active learning, and produced the testing output used for
benchmarking.   Nevertheless, mw-DGP performs nearly as well as TGP, and both
are much better than the others.  Observe that, at least in some cases, mw-DGP
is able to provide better UQ (lower CRPSs) than TGP. We believe this happens
because TGP makes inefficient use of the data when partitioning, whereas
mw-DGP is able to capture the same spirit of axis-aligned nonstationarity
without imposing hard breaks.

\subsubsection*{Satellite drag}

Our final example involves simulations of atmospheric drag for the Gravity
Recovery and Climate Experiment (GRACE) satellite in low-Earth orbit
\citep{mehta2014modeling} via a corpus of one million LHS-based runs provided
by \citet{sun2019emulating}.  These data/simulations involve $p=7$ inputs, and
here we consider randomly sub-sampled GRACE runs in pure Hydrogen of size $(n,
n') = (550, 1000)$.  As the model is not deterministic, fifty training runs
are replicated to provide each model a measure of the uncertainty in the
simulation. The results are shown in the {\em right/third} column of panels of
Figure \ref{f:dgpmets}.  In contrast to our previous examples, there is no
reason to suspect that it might be advantageous to limit nonstationary
modeling to axis-aligned dynamics, monotonic or otherwise, beyond the implicit
regularization that implies.  In fact, observe in the figure that TGP, with
its axis-aligned splits, performs worst of all in some cases.  A DGP is
better than both TGP and an ordinary GP, but in the latter perhaps not by an
impressive margin.  Nevertheless, mw-DGP is by far the best by both metrics.

\section{Discussion}
\label{sec:discuss}

We have provided a new prior on monotonic Gaussian processes (GPs) for a single input
variable that we illustrate is valuable for two disparate classes of problems:
additive monotonic GPs for multiple inputs and as intermediate warpings for deep GPs
(DGPs). There are two key ingredients that make it work well and which distinguish it
from earlier approaches to mono-GP.  One is a reference process that limits
computational effort in the face of cubic computational bottlenecks, and which is
essential for our cumulative-sum-based monotonic transformation.  The other is
elliptical slice sampling (ESS), a powerful but often overlooked MCMC mechanism that
is ideally suited to high-dimensional multivariate normal sampling.  We showed that
our mono-GPs and our monowarped DGPs (mw-DGPs) outperform ordinary GP and DGP
alternatives. In our Appendix we also show that our methods are faster, except
perhaps on the smallest problems.

We note that many other statistical methods exist to model
monotonic response surfaces
\citep{tutz2007genmonotonic,cai2007bayesisotonic,bornkamp2009bayesian},
but the literature on imposing such a constraint for GPs is more limited.
Moreover, our method fits nicely within the context of other frameworks, such
as DGPs. We speculate that other simple constraints on GPs (and DGP
warpings), besides monotonicity, can be handled in a similar way, as long as a
transformation can be used in conjunction with ESS.  This could be
particularly powerful in small-data contexts, e.g., with expensive computer
simulations where much is known about the physics.

As currently implemented, mono-GP does not perform as well when the
response surface is not coordinate-wise additive. [See Appendix
\ref{app:nonadditive}.] Future research could expand mono-GP to multiple inputs by
some construct other than additivity. For example, it may be worth exploring active
subspace-based GPs \citep{wycoff2023} in order to rotate into an appropriate
axis-aligned regime. Additionally, mono-GP could integrate previous work using
Gaussian process projection \citep{lin2014bayesian} that generalizes beyond the class
of additive models.

\subsection*{Acknowledgments}

RBG and SDB are grateful for funding from NSF CMMI 2152679. This work has been
approved for public release under LA-UR-24-28080. SDB, DO and LJB were funded
by Laboratory Directed Research and Development (LDRD) Project 20220107DR.

\bibliography{monoGP}
\bibliographystyle{jasa}

\appendix

\section{Additional methodological details}

Here we provide some auxiliary analysis and implementation notes supporting
method choices described in the main body of our manuscript.

\subsection{Choosing reference grid density}
\label{app:approx}

\begin{figure}[ht!]
\centering
\includegraphics[scale=0.53,trim=25 15 25 50]{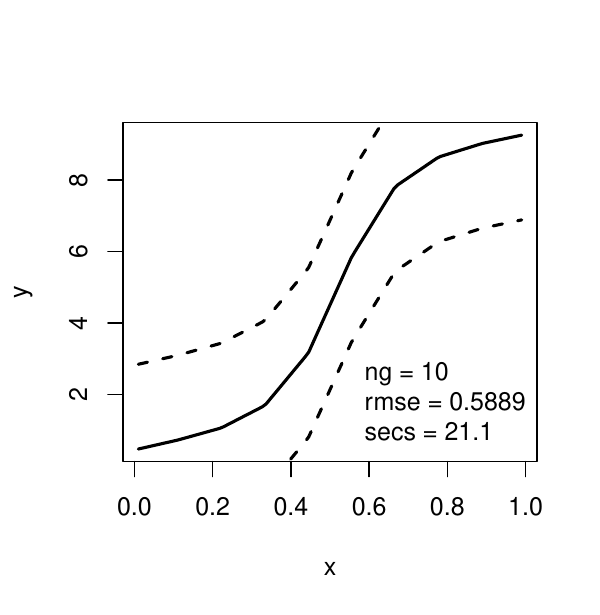}
\includegraphics[scale=0.53,trim=55 15 25 50,clip=TRUE]{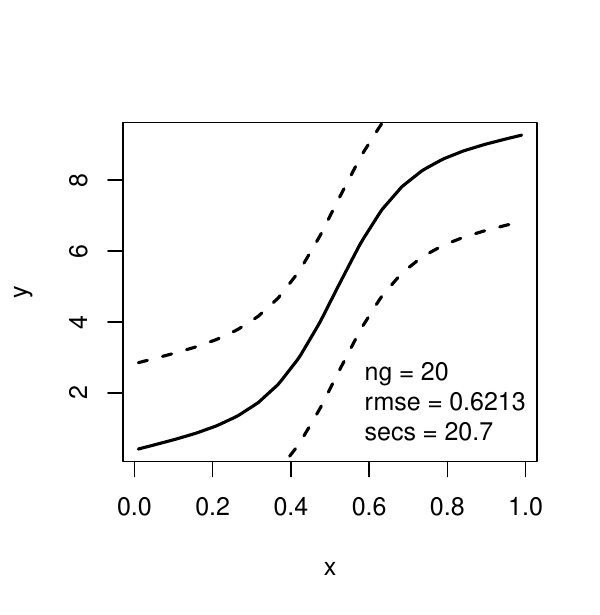}
\includegraphics[scale=0.53,trim=55 15 25 50,clip=TRUE]{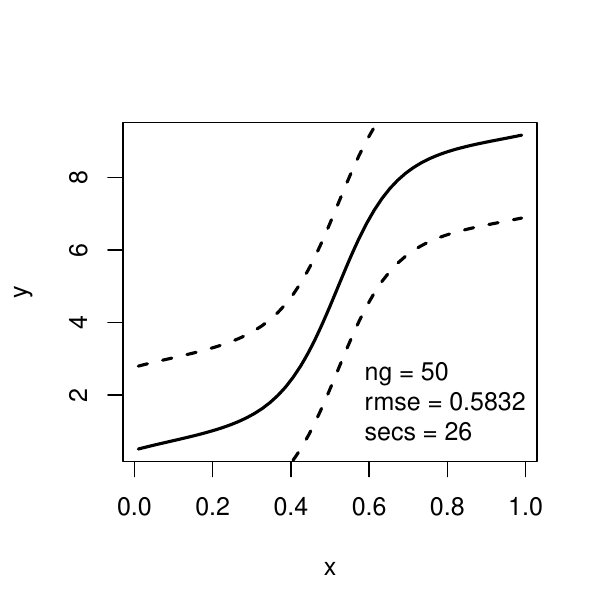}
\includegraphics[scale=0.53,trim=55 15 25 50,clip=TRUE]{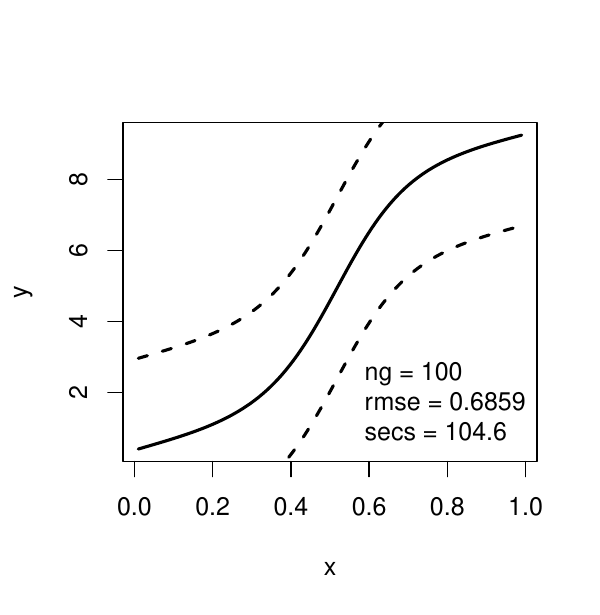}
\caption{1d logistic for varying $n_g$. \label{f:logit1ng}}
\end{figure}

Below we provide two views into accuracy as the size $n_g$ of the reference
grid $X_g$ is increased.  The first is visual, and the second based on
out-of-sample accuracy.  Figure \ref{f:logit1ng} shows views of the 1d
logistic predictive surface as $n_g$ is varied in $\{10, 20, 50, 100\}$.
Otherwise we use the same setup as our illustration in Section \ref{sec:mono1}
with the exception of doubling the number of MCMC iterations in order to
stabilize RMSE calculations.
Observe in the {\em left-most} panel, corresponding to $n_g = 10$, that the
linear interpolation in the mean/black line is visibly detectable.  It may
also be visible in the second panel, corresponding to $n_g = 20$, depending on
screen resolution/magnification or printer quality.  When we zoom in on a
high-quality laptop screen we can see the line segments in the $n_g = 20$
plot.  But this is not the case for either of the {\em right} pair of
panels.

Now consider accuracy out-of-sample, provided in the {\em bottom-right} of
each panel.  This number seems unaffected by $n_g$.  These results come from
just one repetition, so there is MC error which remains unaccounted for, but
hold that thought.  Computing time, also provided in the {\em bottom-right},
is very much affected by $n_g$, especially moving from $n_g = 50$ to $n_g =
100$, with the latter being quite extreme by contrast. So we settled on $n_g =
50$ as a default.  It is worth remarking that $n_g$ also has an impact on the
amount of storage required to retain samples from latent $F_g$ for later use.

Next we pivot to a more exhaustive study on a more challenging,
high-dimensional problem: Lopez--Lopera 5d from Section \ref{sec:add}.  We
vary $n_g$ as above, and track RMSE and computing time.  
\begin{figure}[ht!]
\centering
\includegraphics[scale=0.8,trim=0 20 0 40,clip=TRUE]{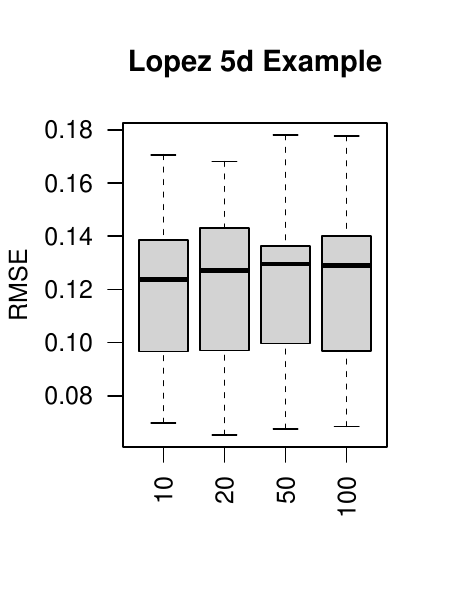}
\includegraphics[scale=0.8,trim=0 20 0 40,clip=TRUE]{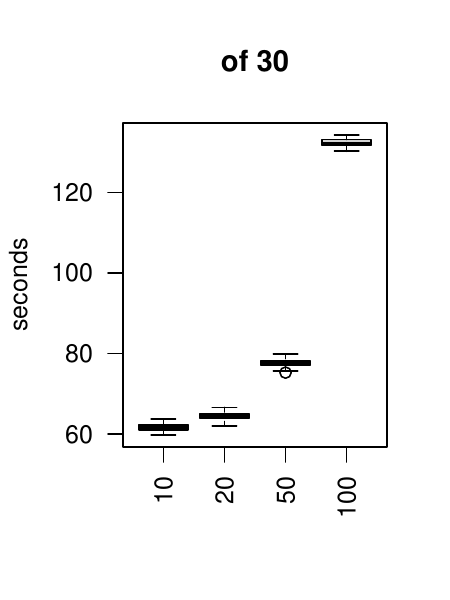}
\vspace{-0.25cm}
\caption{Varying $n_g$ on Lopez--Lopera 5d. \label{f:lopez5ng}}
\end{figure}
Figure \ref{f:lopez5ng} collects the resulting metrics.   It appears that
$n_g$ has no detectable effect on accuracy, but a profound one on execution
time past $n_g = 50$ or so.  It could be that by choosing an even smaller
$n_g$ for our experiments throughout the paper we would have had the same
accuracy and UQ, but a much faster execution time (our comparison in
Appendix \ref{app:timing} might have come out more favorably for small $n$).

We remark that it may be interesting to explore non-uniform
reference grids of varying size, akin to knot placement in adaptive
quadrature. While this could be advantageous, it would require knowledge of
the response surface, presenting a chicken-or-egg problem. We have yet to
encounter any scenarios where a uniform grid of modest size is insufficient.
If a denser grid is warranted in some regions, it would be easiest to increase
the density of the entire grid, which only marginally affects runtime.

\subsection{Variations on the monotonic transformation}
\label{app:prior}

We experimented with several variations on the mono-GP prior, and here we
report comparatively on two which percolated to the top.  The first is the one
summarized in Section \ref{sec:mono1} and described in Eq.~(\ref{eq:mono}).
The second is very similar, but instead of pre-exponentiating, we
subtract off the minimum, i.e, $Z^{(1)} = Z^{(0)} - \min Z^{(0)}$.  This has
the same overall effect of forcing positivity, but it does so linearly.
\begin{figure}[ht!]
\centering
\includegraphics[scale=0.6, trim=0 15 15 50,clip=TRUE]{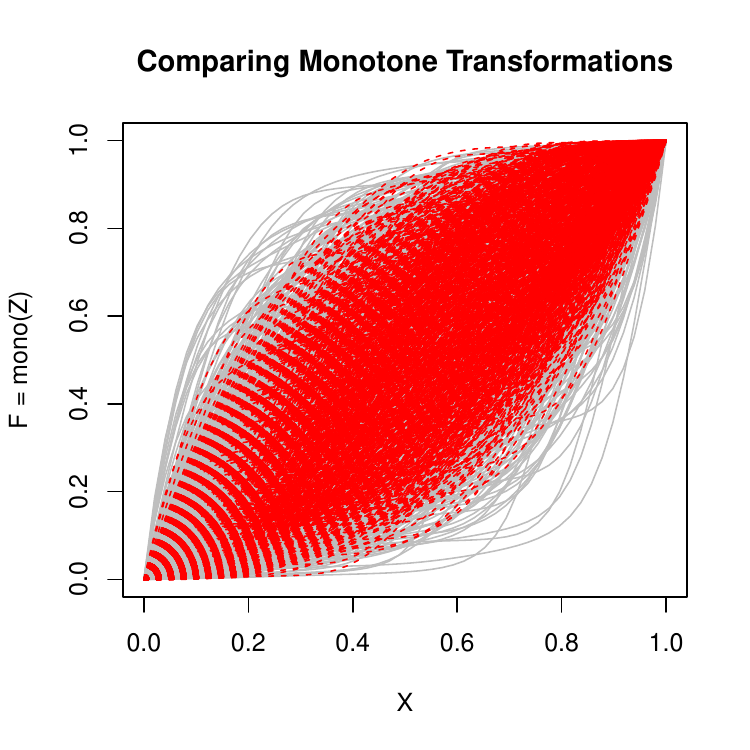}
\includegraphics[scale=0.6, trim=0 15 0 50,clip=TRUE]{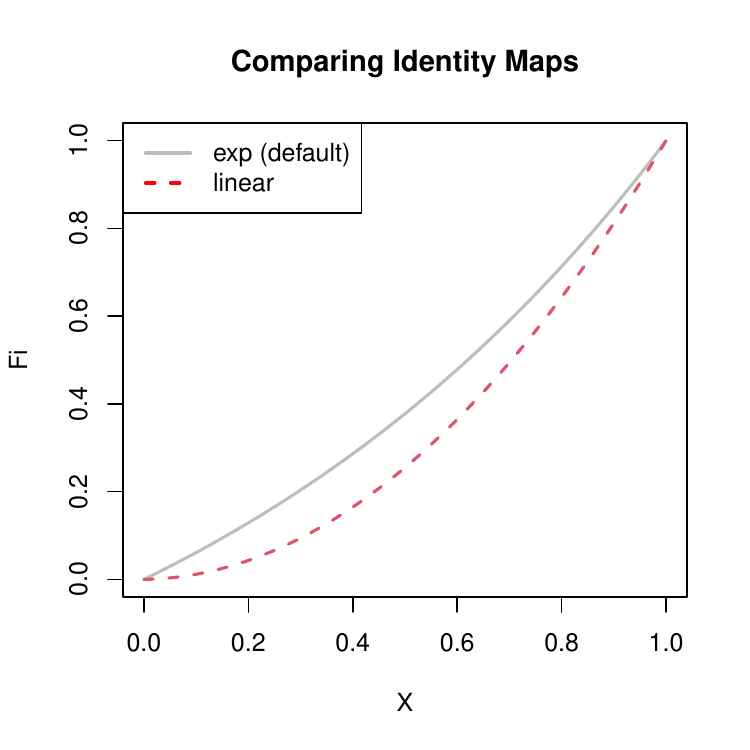}
\caption{Converting 1000 samples from GP prior in two ways (left) and
doing the same thing to the identity map (right). \label{f:prior_lin}}
\end{figure}
Figure \ref{f:prior_lin} provides a comparison between these two priors as
regards 1d monotonic functions.  The {\em left} panel of the figure shows 1000
samples from the prior, in gray for Eq.~(\ref{eq:mono}) and in red dashed
for the new
``linear'' version.  Observe how the spread of gray lines is wider.  We liked
this because it implied a more diffuse prior over random monotonic functions.

On the {\em right} we show how an identity functional relationship maps under
the two transformations.  The identity is important for our DGP warping
application [Section \ref{sec:mwDGP}], where in that context it means that the
$X = W$, i.e., a null warping, implying an ordinary stationary GP -- a
sensible base case \citep{sauer2023active}.  Observe in the {\em right} panel
that both priors add some curvature to the identity, but that it is more
pronounced for the red ``linear'' option, which may at first be
counter-intuitive.  The curvature is coming from the cumulative sum in
Eq.~(\ref{eq:mono}), and by pre-exponentiating we are undoing some of that
curvature.

\subsection{Bespoke linear interpolation}
\label{app:fo_approx}

{\sf R}'s built-in {\tt approx} function provides linear interpolation, but it
has two downsides in our setting.  One is that it doesn't linearly
interpolate beyond the most extreme ``reference'' locations $X_g$.  By
default, it provides {\tt NA} in those regions, or it can provide a constant
interpolation when specifying {\tt rule = 2}.  So any training
data $X_n$ or predictive locations $\mathcal{X}_{n'}$ more extreme than $X_g$
would suffer a loss of resolution at best.  We found this drawback easy to fix
with our own, bespoke implementation.  We simply take the slope and intercept
from the adjacent, within-boundary pair and apply it on the other side of the
boundary.

A second downside is computational.  The {\tt approx} function re-calculates
indices from the interpolating set (say $X_n$) to the reference ($X_g$) each
time it is called, say in each MCMC iteration.  But our $X_n$ and $X_g$ are
fixed throughout the MCMC, so this effort is redundant thousands of times
over.  If those indices could be pre-calculated, and passed from one call to
the next, there could be potentially substantial speedups. Our implementation,
which may be found in our supplementary material and Git repository, provides such a
pre-indexing calculation, which is then passed along as needed.  We call this
(including our linear extrapolation above) a ``fixed-order'' approximation (or
``fo'' for short), and the functions are called  \verb!fo_approx_init! and
\verb!fo_approx! respectively.  We have found that it can yield a 2-3$\times$
speedup over ordinary {\tt approx}. However, linear interpolation is not the
only operation involved in our MCMC. Figure \ref{f:timing}, coming later in
Appendix \ref{app:timing}, shows that the speedup is closer to 0.5$\times$,
however the gap widens as the training data size $n$ is increased.

Note that {\sf R} also provides a function called {\tt approxfun} which, at
first glance, appears to offer a similar pre-processing based efficiency.
However, it is designed for fixed input/output ($x$ and $y$-values) not fixed
input and predictive values ($x$ and $x'$ say), which is what we have.  We
have novel $y$-values, coming from from an MVN via ESS in each MCMC iteration,
so {\tt approxfun} does not offer us any speedups.

\section{Additional empirical results}

Here we provide two additional sets of empirical results, one for mono-GP and
one for mw-DGP, along with the same comparators we entertained in the main
text for examples from these two classes of problems.  Then we turn to a
timing comparison for both methods.

\subsection{Logistic mono-GP comparison}
\label{app:logistic}

\begin{figure}[ht!]
\centering
\includegraphics[scale=0.80, trim=0 5 0 50,clip=TRUE]{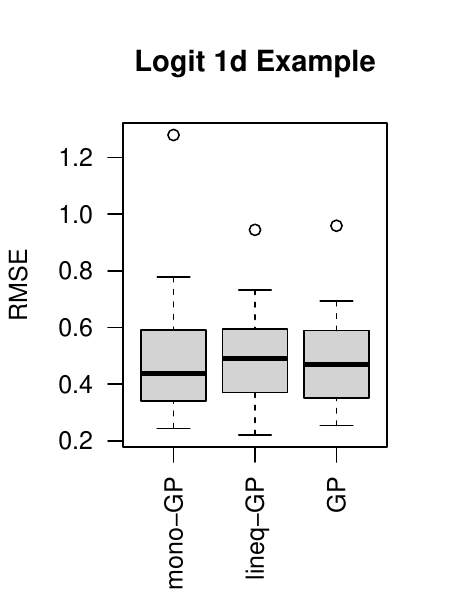}
\includegraphics[scale=0.80, trim=0 5 10 50,clip=TRUE]{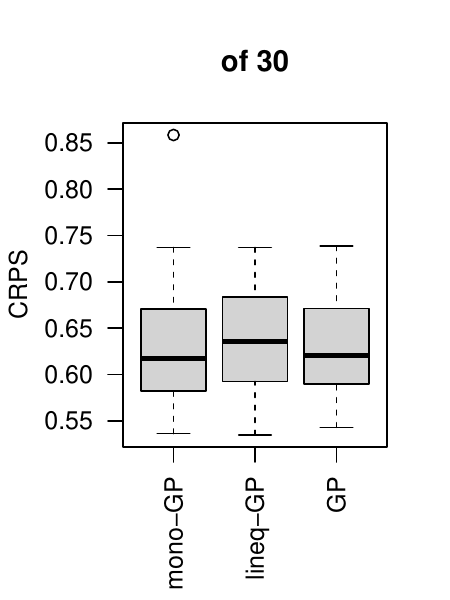}
\vspace{-0.1cm}
\caption{Metrics on the 1d logit MC experiment.  Smaller is better for all.
\label{f:logit1bench}}
\end{figure}

Here we report on a MC experiment in the style of Section \ref{sec:monobench}
but on the 1d logistic data used for illustrations in Section \ref{sec:mono1}.
In this simple setting, with $n=20$ training data locations, there is more
than enough information in the data for almost any nonlinear regression to
work well.  Consequently, results provided in Figure \ref{f:logit1bench}
indicate that all three methods are more-or-less equally good.
In particular, although the ordinary GP cannot guarantee monotonicity, as
shown in Figure \ref{f:logit1}, it nevertheless provides accurate predictions
with good UQ.  In fact, since it integrates over the latent field analytically
its metrics have lower MC error (narrower boxplots).  Mono-GP and lineq-GP
both require MC integration, the accuracy of which is determined by a limited
amount of posterior sampling.

\subsection{Michalewicz deep GP comparison}
\label{app:michael}

Here we report on a MC experiment in the style of Section \ref{sec:gdpbench}.
Consider the ``Michalewicz'' function \citep[VLSE;][]{surjanovic2013virtual}:
$$
f(x) = - \sum_{i=1}^p \sin(x_i) \sin^{2m}\left(\frac{i x_i^2 }{\pi}\right), 
\quad \quad x \in [0, \pi]^p.
$$
For variety, we ran our MC experiment in 3d ($p=3$) and set $m = 10$, which is
the recommended value.  Inputs were coded to the unit cube and
back-transformed as needed. Training/testing sets are generated from an LHS
design with $(n, n') = (100, 1000)$.
\begin{figure}[ht!]
\centering
\includegraphics[scale=0.8,trim=0 8 20 55,clip=TRUE]{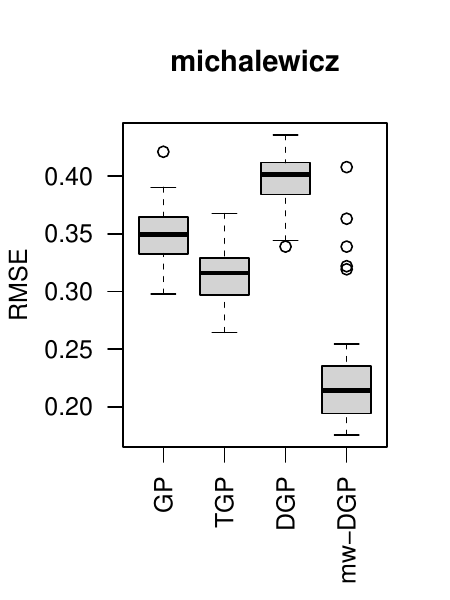}
\includegraphics[scale=0.8,trim=0 8 20 55,clip=TRUE]{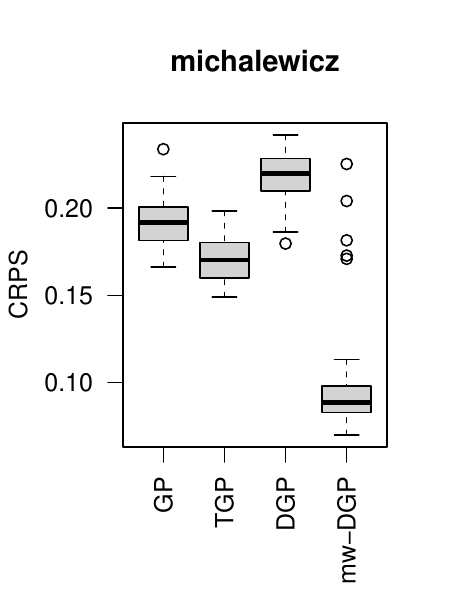}
\vspace{-0.1cm }
\caption{Michalewicz (3D) comparison. \label{f:michael}}
\end{figure}
Otherwise the setup is the same as our earlier DGP-based experiments.  Results
are shown in Figure \ref{f:michael}. It is clear that mw-DGP outperforms the
three other methods in terms of both accuracy and UQ. Interestingly, the
ordinary GP does better than DGP. The extreme flexibility of a DGP may in fact
may be a detriment to its performance without suitable regularization, e.g.,
as provided by mw-DGP.  However, it is surely possible to engineer an example
which is rotated along a diagonal that would thwart a purely axis-aligned
process.

\subsection{Timing comparisons}
\label{app:timing}

Here we measure fitting time on the 2d Lopez--Lopera [Section
\ref{sec:monobench}] and 3d Michalewicz [Appendix \ref{app:michael}] examples.
Besides varying the training and testing set sizes commensurately ($n = n'$),
there are no other changes from those experiments. Prediction time is not
reported because the additional expense represents less than 15\% of
computational time for all methods (and less than 1\% for our method).
We vary $n$ on an exponential schedule beginning with $n\approx 50$ and
ending with $n\approx 8000$ for 2d Lopex--Lopera and $n\approx 2500$ for
3d Michalewicz.

\begin{figure}[ht!]
\centering
\includegraphics[scale=0.65, trim=0 15 15 45,clip=TRUE]{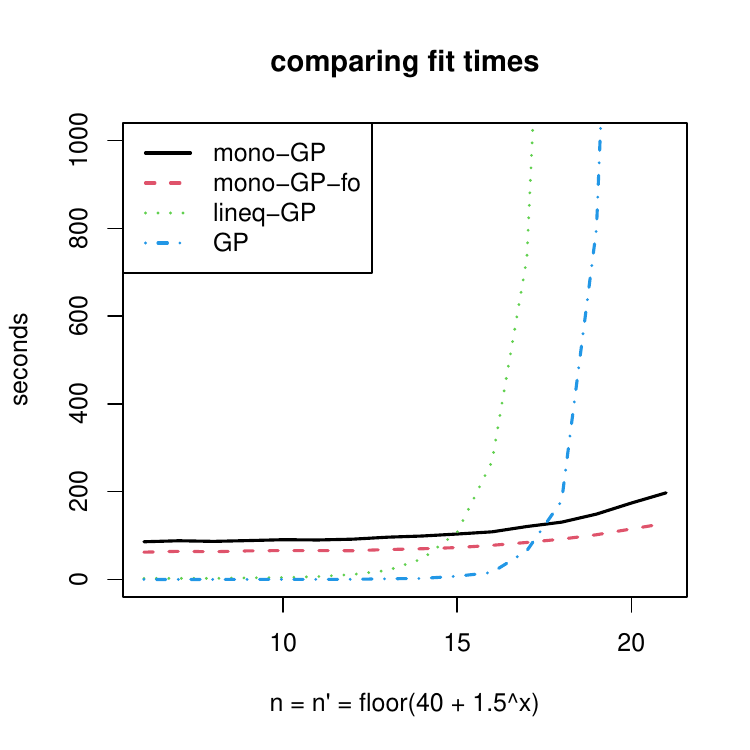}
\includegraphics[scale=0.65, trim=0 15 15 45,clip=TRUE]{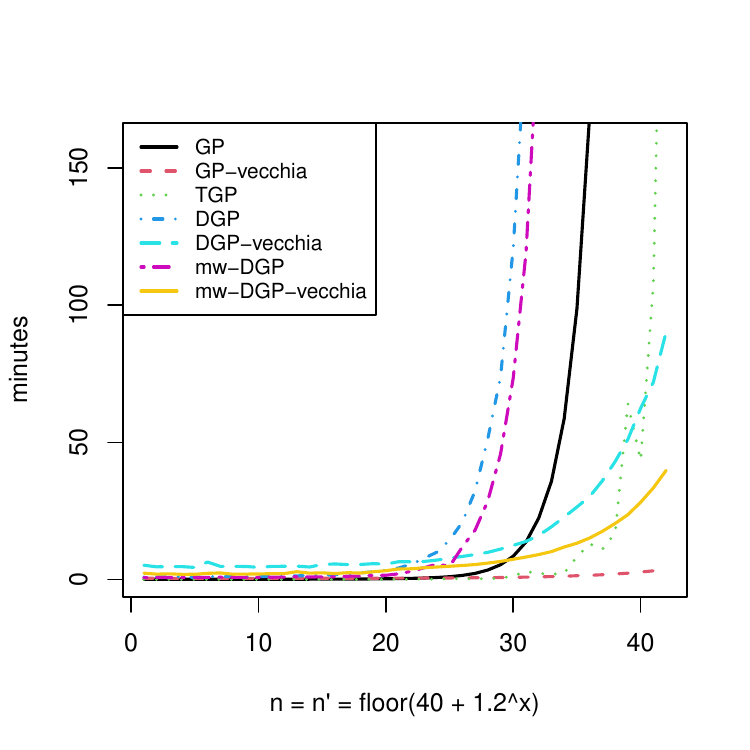}
\caption{Compute times varying $n = n'$ on 5d Lopez--Lopera (left) and 3d
Michalewicz (right) examples from earlier. \label{f:timing}}
\end{figure}

Figure \ref{f:timing} shows the results.  The $y$-axis on both plots is cut
off to enhance visibility for the smaller-$n$ experiments.  Focusing on the
{\em left} panel first, notice that mono-GP is slower (both variations),
relatively speaking, for these.  But eventually the cubic costs of lineq-GP
and the ordinary GP kick in, and those methods become prohibitively slow.
Mono-GP and lineq-GP are similar on computation time for $n\approx 500$, with
the ordinary GP being faster in that instance which we attribute to its more
streamlined {\sf C} implementation.  The break even point for mono-GP and the
ordinary GP is $n\approx 1000$.  When $n\approx 3300$ mono-GP only takes 142
seconds, which is barely double the 71 seconds required for $n=31$.  However,
lineq-GP and the ordinary GP take 30000 and 2000 seconds respectively.  
%This is the reason that we trimmed the $y$-axis in the figure.

The variation labeled ``mono-GP'' uses {\sf R}'s built-in {\tt approx},
whereas ``mono-GP-fo'' uses the thriftier \verb!fo_approx!.  In other experiments
not reported on here, we found that our bespoke \verb!fo_approx! was 2--3x faster
by pre-computing an appropriate indexing, as opposed to re-computing it each
time it is needed. When situated within our mono-GP MCMC, which of course
involves many other calculations, the speedup (shown in the figure) is a more
modest but noticeable 50\%.  We use this ``mono-GP-fo'' as our default (i.e.,
for mono-GP) in the rest of the empirical work reported on in this paper.

On the 3d Michalewicz problem, some methods (e.g. DGP, mw-DGP) took hours, if
not days, to complete for the larger training data sets without use of the
Vecchia approximation. This is illustrated by the $y$-axis only extending to
$\approx 2.5$ hours in the {\em right} panel of Figure \ref{f:timing}. After
$n > 200$, there's essentially no contest in terms of efficiency. But the use
of a fixed grid size in mw-DGP also adds a speedup. At every training set
size, mw-DGP runs faster than DGP. It's the same story when the Vecchia
approximation is employed for both methods. Perhaps more importantly, as the
training set reaches sizes that are orders of magnitude larger, the gap
between mw-DGP-vecchia and DGP-vecchia actually increases. This is because all
operations on the warping layer in mw-DGP-vecchia use the fixed size grid,
thereby locking in the cubic costs at $\mathcal{O}(n_g^3)$. DGP-vecchia
approximates the warping layer, so the computational costs still increase
linearly as training set size grows.

TGP's execution time is a little more variable, depending on the number of
partitions it creates. Although it is able to handle larger training data
than DGP, with execution time not spiking until $n\approx 700$, TGP still
cannot match the efficiency of mw-DGP-vecchia for data of substantial size.
It's no surprise that GP-vecchia, without the addition of a warping layer
(and the extra computation that accompanies it), executes in a smaller time
frame than mw-DGP-vecchia. In fact, it is impossible for mw-DGP-vecchia to
beat GP-vecchia, because they both perform the same approximation on the outer
layer. But with that extra warping layer, mw-DGP-vecchia easily wins in RMSE
and CRPS, as shown in \ref{f:michael}.

\subsection{Additional sensitivity visuals}
\label{app:sens}

\begin{figure}[ht!]
\centering
\includegraphics[scale=0.6, trim=30 60 30 50,clip=TRUE]{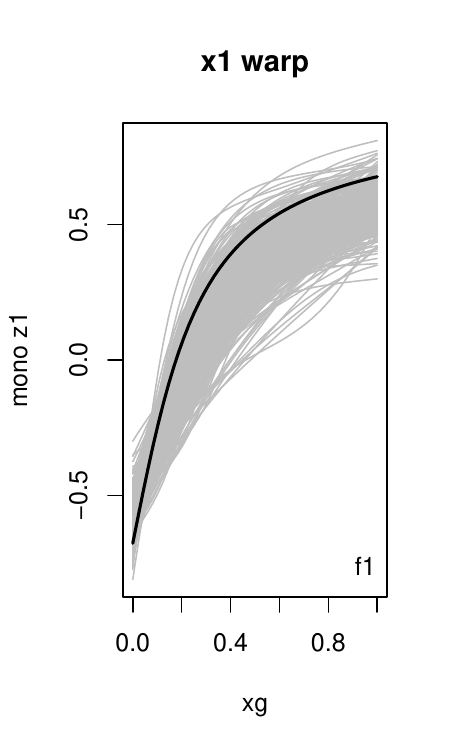}
\includegraphics[scale=0.6, trim=55 60 30 50,clip=TRUE]{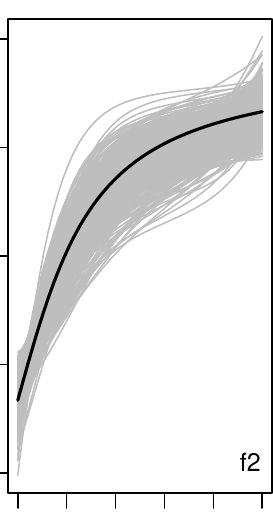}
\includegraphics[scale=0.6, trim=55 60 30 50,clip=TRUE]{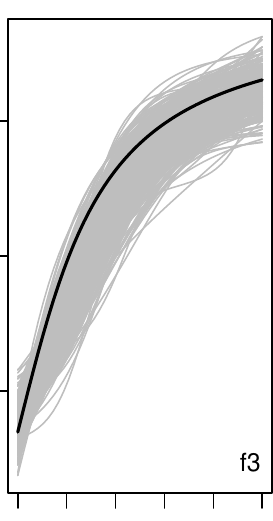}
\includegraphics[scale=0.6, trim=55 60 30 50,clip=TRUE]{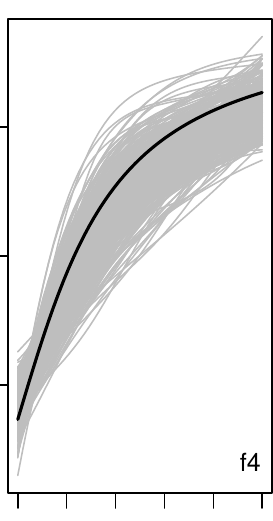}
\includegraphics[scale=0.6, trim=55 60 30 50,clip=TRUE]{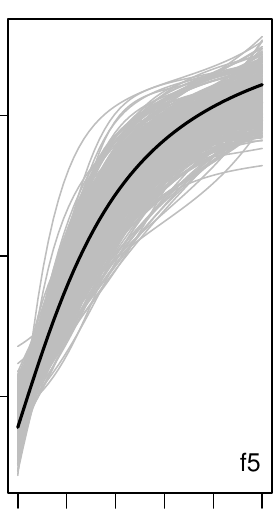}
\includegraphics[scale=0.6, trim=30 15 30 50,clip=TRUE]{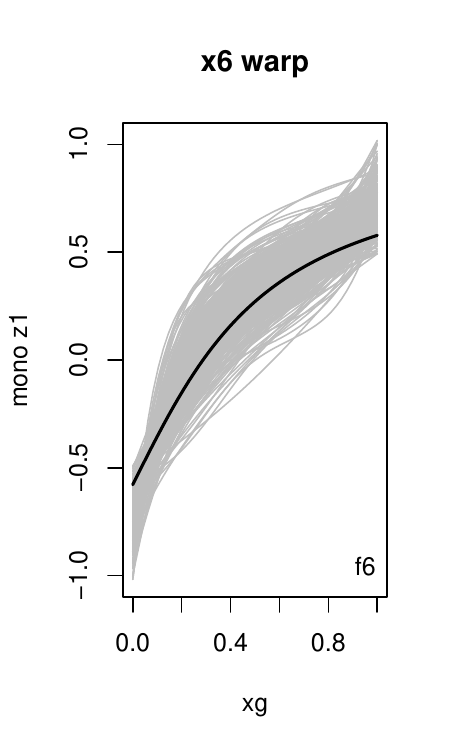}
\includegraphics[scale=0.6, trim=55 15 30 50,clip=TRUE]{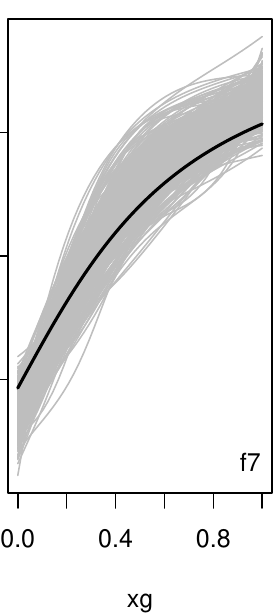}
\includegraphics[scale=0.6, trim=55 15 30 50,clip=TRUE]{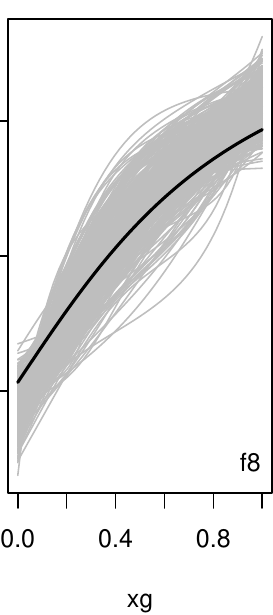}
\includegraphics[scale=0.6, trim=55 15 30 50,clip=TRUE]{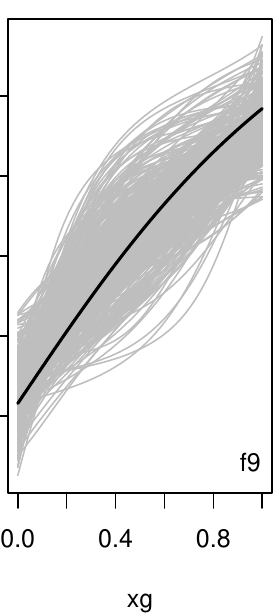}
\includegraphics[scale=0.6, trim=55 15 30 50,clip=TRUE]{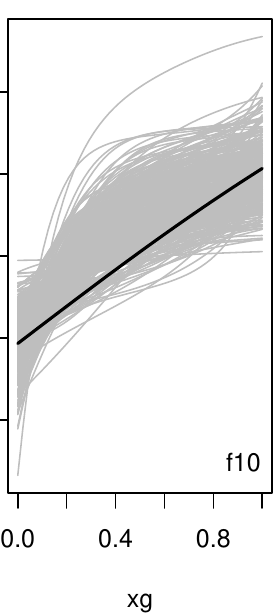}
\caption{Latent functions for arctan example compared to the truth. \label{f:arctan10dlatent}}
\end{figure}

\begin{figure}[ht!]
\centering
\includegraphics[scale=0.65, trim=0 15 15 45, clip=TRUE]{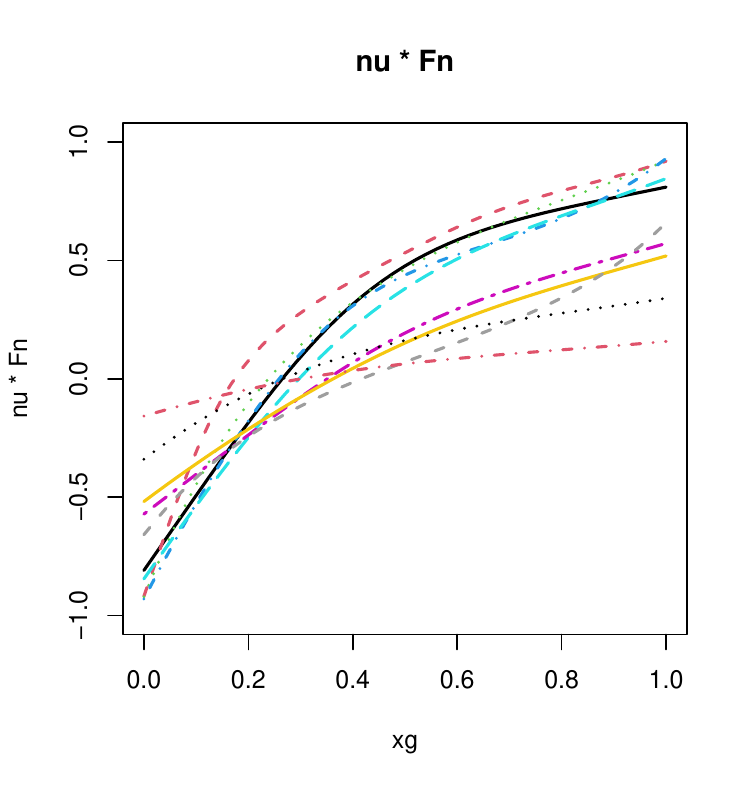}
\caption{Sensitivity via posterior means of $\nu_j
F_n^{\cdot j}$ on 10d Lopez--Lopera artcan example.%Each line corresponds to
%one of $j=1, \dots, p$. 
\label{f:sens}}
\end{figure}

We omitted the Lopez--Lopera arctan sensitivity plots in Section
\ref{sec:monobench} to save space, so here they are in Figure
\ref{f:arctan10dlatent}.  Each panel shows all saved posterior samples of
$F_n^{\cdot, j}$ for $j=1,\dots, p = 10$.
Observe that they are all quite similar in shape.  In this example, the
$\nu$-vector is key in weighting contributions from each individual input.  In
Figure \ref{f:sens} we show the posterior means of samples of $\nu^{(t)}
\times F_n^{(t)}$, where each line corresponds to one of the columns of this
matrix.
Observe that some of those lines span a wider swath of the $y$-axis, amplifying
their contribution to the overall predictive mean through the additive model.

\subsection{Non-additive monotonic response surfaces}
\label{app:nonadditive}

\begin{figure}[ht!]
\centering
\includegraphics[scale=0.8,trim=0 4 30 50,clip=TRUE]{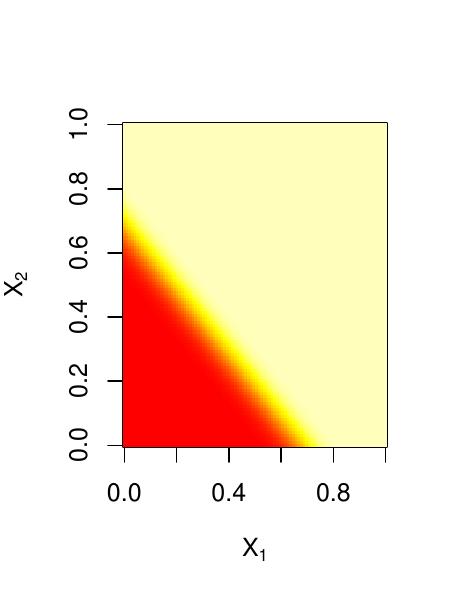}
\includegraphics[scale=0.8,trim=0 4 30 50,clip=TRUE]{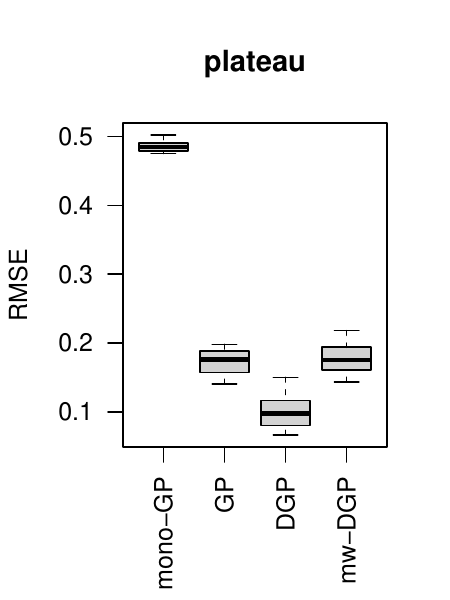}
\includegraphics[scale=0.8,trim=0 0 30 50,clip=TRUE]{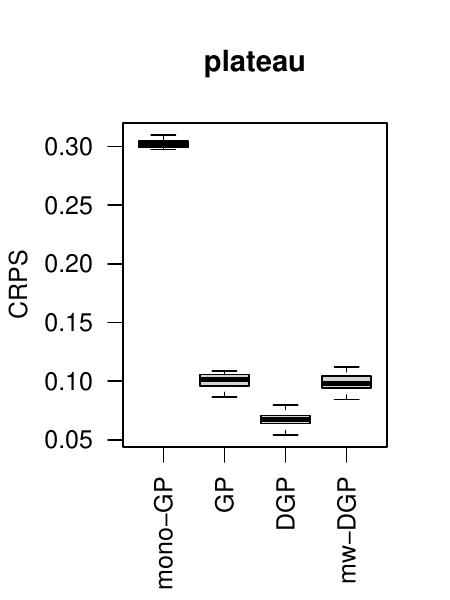}
\vspace{-0.1cm }
\caption{Plateau function (left); Performance metrics on 30 training/test sets of
size $n=100$ and $n'=1600$ (middle, right). \label{f:plateau}}
\end{figure}

We have demonstrated that our method outperforms its competitors when the
response surface is constructed additively, as the title of our paper
suggests. However, at the direction of a referee we explore here an expanded
family of functions that adhere to monotonicity in each dimension, but are not
necessarily additive. In short, we wish to explore how well mono-GP
performs when the model is mis-specified. Consider the ``plateau'' function
\begin{equation}\label{eq:plateau}
f(x) = 2 \cdot \Phi\left(\sqrt{2} \left(-4 - 3\sum_{i=1}^d x_i\right)\right) - 1
\quad\textrm{for}\; x\in[-2, 2]^d \; \left(\textrm{scaled to}\; x\in[0, 1]^d
\right),
\end{equation}
as displayed in the left panel of Figure \ref{f:plateau}. Note that the
response is monotonically increasing in both $x_1$ and $x_2$, but it's clear
from Eq.~(\ref{eq:plateau}) that the function is not coordinate-wise additive.
More succinctly, it is monotone along the $d$-dimensional diagonal. To assess
performance of our method in this context, we conduct a MC experiment
comparing mono-GP to an ordinary GP, ordinary DGP, and mw-DGP on 30 unique
training/testing sets with $(n, n') = (100, 1600)$. Results are shown in the
middle and right panels of Figure \ref{f:plateau}.

The ordinary DGP wins both in terms of predictive accuracy and UQ.
Interestingly, mw-DGP performs similarly to the ordinary GP, indicating that
the ordinary DGP relies on relationships across dimensions to appropriately
warp the inputs. By a clear margin, mono-GP falls behind the other three. Our
proposed method, therefore, does not appear to excel in generality. While we
note that this example is as pathological as one can get for a monotonic,
non-additive surface, it illustrates that mono-GP needs modification to
account for non-axis-aligned changes in the response surface or some type of
pre-processing step to rotate the space into an axis-aligned format. This is a
subject for future work.

\end{document}